\documentclass[conference, 10pt]{IEEEtran}
\usepackage[dvips]{graphicx}
\usepackage{amssymb,amsmath}
\usepackage{subfig}
\usepackage{cite}
\usepackage{colortbl}
\usepackage{url}

\IEEEoverridecommandlockouts
\newcommand{\n}[1]{\mathcal{N}_{#1}}
\newcommand{\thru}[1]{\mathcal{T}_{#1}}
\newcommand{\band}{\mathcal{B}}

\begin{document}

\title{Modeling Network Coded TCP: \\Analysis of Throughput and Energy Cost
}
\author{
\authorblockN{MinJi Kim\authorrefmark{1}, Thierry Klein\authorrefmark{2}, Emina Soljanin\authorrefmark{2}, Jo\~{a}o Barros\authorrefmark{4}, Muriel M\'{e}dard\authorrefmark{1}\vspace*{-0.3cm}}
\thanks{\authorrefmark{1}M. Kim and M. M\'edard are with the RLE at the Massachusetts Institute of Technology, MA USA (e-mail: \{minjikim, medard\}@mit.edu).}
\thanks{\authorrefmark{2}T. Klein and E. Soljanin are with the Alcatel-Lucent Bell Laboratories, NJ USA (e-mail: thierry.klein@alcatel-lucent.com, emina@research.bell-labs.com).}
\thanks{\authorrefmark{4}J. Barros is with the Department of Electrical and Computer Engineering at the University of Porto, Portugal (e-mail: jbarros@fe.up.pt).}
\thanks{This work is supported by the MIT-Portugal Program under award No: 014098-153.}
}

\maketitle

\begin{abstract}
We analyze the performance of TCP and TCP with network coding (TCP/NC) in lossy networks. We build upon the framework introduced by Padhye et al. and characterize the throughput behavior of classical TCP and TCP/NC as a function of erasure probability, round-trip time, maximum window size, and duration of the connection. Our analytical results show that network coding masks random erasures from TCP, thus preventing TCP's performance degradation in lossy networks. It is further seen that TCP/NC has significant throughput gains over TCP.

In addition, we show that TCP/NC may lead to cost reduction for wireless network providers while maintaining a certain quality of service to their users. We measure the cost in terms of number of base stations, which is highly correlated to the energy, capital, and operational costs of a network provider. We show that increasing the available bandwidth may not necessarily lead to increase in throughput, particularly in lossy networks in which TCP does not perform well. We show that using protocols such as TCP/NC, which are more resilient to erasures, may lead to a throughput commensurate the bandwidth dedicated to each user.
\end{abstract}

\section{Introduction}\label{sec:introduction}

The Transmission Control Protocol (TCP) is one of the core protocols of today's Internet Protocol Suite. TCP was designed for reliable transmission over wired networks, in which losses are generally indication of congestion. This is not the case in wireless networks, where losses are often due to fading, interference, and other physical phenomena. Consequently, TCP's performance in wireless networks is poor when compared to the wired counterparts as shown e.g. in \cite{caceres,TCP_Kurose}. There has been extensive research to combat these harmful effects of erasures and failures \cite{sack, stcp,Tian05tcpin}; however, TCP even with modifications does not achieve significant improvement. For example, there has been suggestions to allow TCP sender to maintain a large transmission window to overcome the random losses within the network. However, as we shall show in this paper, just keeping the window open does not lead to improvements in TCP's performance. Even if the transmission window is kept open, the sender can not transmit additional packets into the network without receiving acknowledgments. References \cite{hari,Tian05tcpin} give an overview and a comparison of various TCP versions over wireless links.

Some relief may come from network coding \cite{ahlswede}, which has been introduced as a potential paradigm to operate communication networks, in particular wireless networks. Network coding allows and encourages mixing of data at intermediate nodes, which has been shown to increase throughput and robustness against failures and erasures \cite{algebraic}. There are several practical protocols that take advantage of network coding in wireless networks \cite{xor,more,codeOr,costa}. In order to combine the benefits of TCP and network coding, \cite{tcpnc} proposes a new protocol called TCP/NC. 

In this paper, we present a performance evaluation of TCP as well as TCP/NC in lossy networks. We adopt and extend the TCP model in \cite{TCP_Kurose} -- i.e. we consider standard TCP with Go-Back-N pipe lining. Thus, the standard TCP discards packets that are out-of-order. We analytically show the throughput gains of TCP/NC over standard TCP. We characterize the steady state throughput behavior of both TCP and TCP/NC as a function of erasure rate, round-trip time (RTT), and maximum window size. Furthermore, we use NS-2 (Network Simulator \cite{ns}) to verify our analytical results for TCP and TCP/NC. Our analysis and simulations show that TCP/NC is robust against erasures and failures. TCP/NC is not only able to increase its window size faster but also maintain a large window size despite losses within the network. Thus, TCP/NC is well suited for reliable communication in lossy networks. In contrast, standard TCP experiences window closing as losses are mistaken to be congestion.

We use the model for TCP/NC's and TCP's performance to study their effect on cost of operating a network. In particular, we show that maintaining or even improving users' quality of experience may be achieved without installing additional network infrastructure, e.g. base stations. We measure users' quality of experience using the throughput perceived by the user. We make a clear distinction between the terms throughput and \emph{bandwidth}, where throughput is the number of \emph{useful} bits over unit time received by the user and bandwidth is the number of bits transmitted by the base station per unit time. In essence, bandwidth is indicative of the resources provisioned by the service providers; while throughput is indicative of the user's quality of experience. For example, the base station, taking into account the error correction codes, may be transmitting bits at 10 megabits per second (Mbps), i.e. bandwidth is 10 Mbps. However, the user may only receive useful information at 5 Mbps, i.e. throughput is 5 Mbps.

The disparity between throughput and bandwidth used can be reduced by using a transport protocol that is more resilient to losses. One method is to use multiple base stations simultaneously (using multiple TCP connections \cite{parallel} or multipath TPC \cite{mptcp2}). However, the management of the multiple streams or paths may be difficult, especially in lossy networks. Furthermore, each path or TCP stream may still suffer from performance degradation in lossy environments \cite{parallel, mptcp2}. We show that erasure-resilient protocols such as TCP/NC \cite{tcpnc,analysis} can effectively reduce the disparity between throughput and bandwidth. 

There has been extensive research on modeling and analyzing TCP's performance \cite{analsys_low,analsys_low2,analysis_altman,analysis_baccelli,analysis_garetto,analysis_liu}. Our goal is to present an analysis for TCP/NC, and to provide a comparison of TCP and TCP/NC in a lossy wireless environment. We adopt Padhye et al.'s model \cite{TCP_Kurose} as their model provides a simple yet good model to predict the performance of TCP. It would be interesting to extend and analyze TCP/NC in other TCP models in the literature.

This paper is based on the work from \cite{analysis,energy_analysis}. The paper is organized as follows. In Section \ref{sec:overview}, we provide a brief overview of TCP/NC. In Section \ref{sec:model}, we introduce our communication model. Then, we provide throughput analysis for TCP and TCP/NC in Sections \ref{sec:td} and \ref{sec:tcpnc}, respectively.
In Section \ref{sec:simulations}, we provide simulation results to verify our analytical results in Sections \ref{sec:td} and \ref{sec:tcpnc}. In Section \ref{sec:bs_model}, we present our model for analyzing the cost associated with operating the network, and analyze the number of base stations needed in Section \ref{sec:bs_analysis}. We study the best case scenario in Section \ref{sec:bestcase}, and compare this idealized scenario with those of TCP and TCP/NC in Section \ref{sec:bs_tcpnc}. Finally, we conclude in Section \ref{sec:conclusions}.

\section{Overview of TCP/NC}\label{sec:overview}

\begin{figure}[tbp]
\begin{center}\vspace*{.4cm}
\includegraphics[width=.35\textwidth]{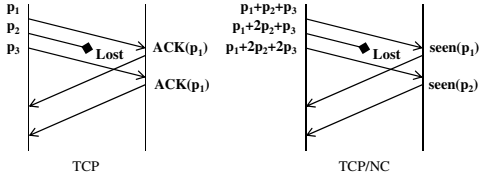}
\end{center}\vspace*{-.2cm}\caption{Example of TCP and TCP/NC. In the case of TCP, the TCP sender receives duplicate ACKs for packet $\mathbf{p_1}$, which may wrongly indicate congestion. However, for TCP/NC, the TCP sender receives ACKs for packets $\mathbf{p_1}$ and $\mathbf{p_2}$; thus, the TCP sender perceives a longer round-trip time (RTT) but does not mistake the loss to be congestion.}\label{fig:example}\vspace*{-.2cm}
\end{figure}

Reference \cite{tcpnc} introduces a new \emph{network coding} layer between the TCP and IP in the protocol stack. The network coding layer intercepts and modifies TCP's acknowledgment (ACK) scheme such that random erasures does not affect the transport layer's performance. To do so, the \emph{encoder}, the network coding unit under the sender TCP, transmits $R$ random linear combinations of the buffered packets for every transmitted packet from TCP sender. The parameter $R$ is the \emph{redundancy factor}. Redundancy factor helps TCP/NC to recover from random losses; however, it cannot mask correlated losses, which are usually due to congestion. The \emph{decoder}, the network coding unit under the receiver TCP, acknowledges \emph{degrees of freedom} instead of individual packets, as shown in Figure \ref{fig:example}. 
Once enough degrees of freedoms are received at the decoder, the decoder solves the set of linear equations to decode the original data transmitted by the TCP sender, and delivers the data to the TCP receiver.

We briefly note the overhead associated with network coding. The main overhead associated with network coding can be considered in two parts: 1) the coding vector (or coefficients) that has to be included in the header; 2) the encoding/decoding complexity. For receiver to decode a network coded packet, the packet needs to indicate the coding coefficients used to generate the linear combination of the original data packets. The overhead associated with the coefficients depend on the field size used for coding as well as the number of original packets combined. It has been shown that even a very small field size of $\mathbf{F}_{256}$ (i.e. 8 bits = 1 byte per coefficient) can provide a good performance\cite{tcpnc,tcpnc2}. Therefore, even if we combine 50 original packets, the coding coefficients amount to 50 bytes over all. Note that a packet is typically around 1500 bytes. Therefore, the overhead associated with coding vector is not substantial. The second overhead associated with network coding is the encoding and decoding complexity, and the delay associated with the coding operations. Note that to affect TCP's performance, the decoding/encoding operations must take substantial amount of time to affect the round-trip time estimate of the TCP sender and receiver. However, we note that the delay caused the coding operations is negligible compared to the network round-trip time. For example, the network round-trip time is often in milliseconds (if not in hundreds of milliseconds), while a encoding/decoding operations involve a matrix multiplication/inversion in $\mathbf{F}_{256}$ which can be performed in a few microseconds. 


\section{A Model for Congestion Control}\label{sec:model}

We focus on TCP's congestion avoidance mechanism, where the congestion control window size $W$ is incremented by $1/W$ each time an ACK is received. Thus, when every packet in the congestion control window is ACKed, the window size $W$ is increased to $W+1$. On the other hand, the window size $W$ is reduced whenever an erasure/congestion is detected.

We model TCP's behavior in terms of \emph{rounds} \cite{TCP_Kurose}. We denote $W_i$ to be the size of TCP's congestion control window size at the beginning of round $i$. The sender transmit $W_i$ packets in its congestion window at the start of round $i$, and once all $W_i$ packets have been sent, it defers transmitting any other packets until at least one ACK for the $W_i$ packets are received. The ACK reception ends the current round, and starts round $i+1$.

For simplicity, we assume that the duration of each round is equal to a round trip time ($RTT$), independent of $W_i$. This assumes that the time needed to transmit a packet is much smaller than the round trip time. This implies the following sequence of events for each round $i$: first, $W_i$ packets are transmitted. Some packets may be lost.
The receiver transmits ACKs for the received packets. (Note that TCP uses cumulative ACKs. Therefore, if the packets $1, 2, 3, 5, 6$ arrive at the receiver in sequence, then the receiver ACKs packets $1, 2, 3, 3, 3$. This signals that it has not yet received packet 4.) Some of the ACKs may also be lost. Once the sender receives the ACKs, it updates its window size. Assume that $a_i$ packets are acknowledged in round $i$. Then, $W_{i+1} \leftarrow W_i + a_i/W_i$.

\begin{figure*}[tbp]
\begin{center}
\subfloat[TCP]{\label{fig:td-tcp}\includegraphics[width=.35\textwidth]{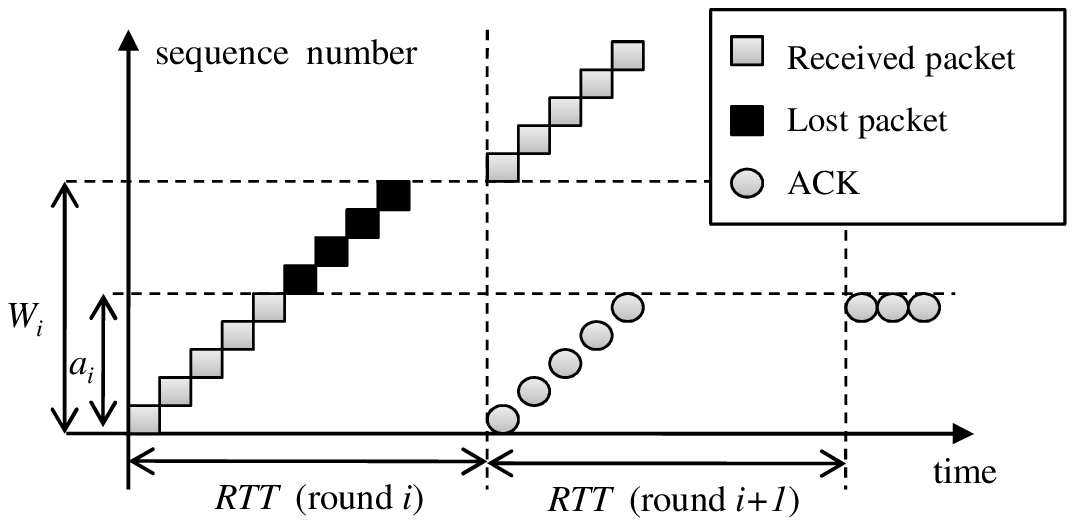}}
\hspace*{.7cm}
\subfloat[TCP/NC]{\label{fig:td-tcpnc}\includegraphics[width=.35\textwidth]{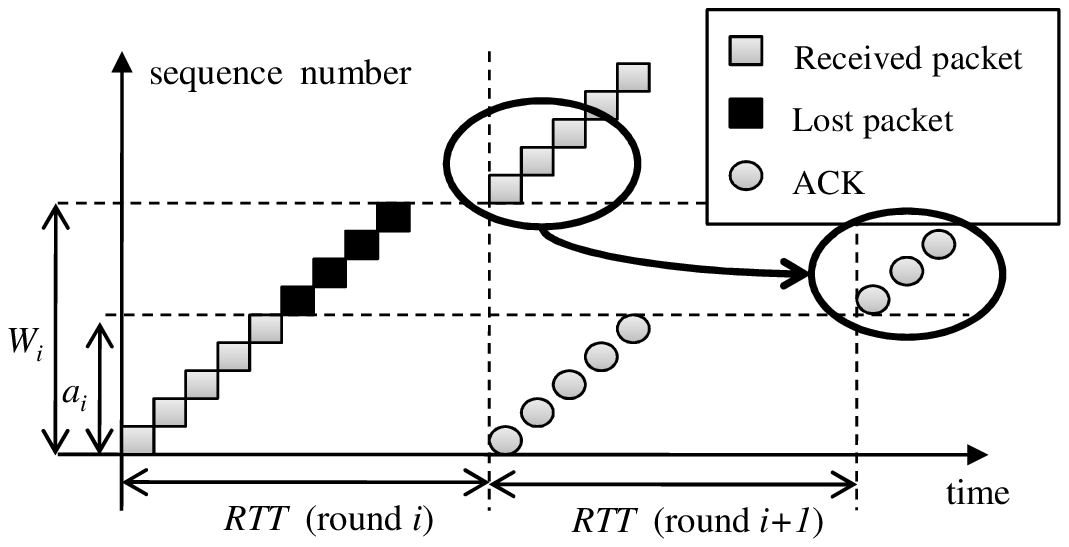}}
\end{center}\vspace*{-.2cm}
\caption{The effect of erasures: TCP experiences triple-duplicate ACKs, and results in $W_{i+2} \leftarrow W_{i+1}/2$. However, TCP/NC masks the erasures using network coding, which allows TCP to advance its window. This figure depicts the sender's perspective, therefore, it indicates the time at which the sender transmits the packet or receives the ACK.}\label{fig:td}\vspace*{-.2cm}
\end{figure*}

TCP reduces the window size for congestion control using the following two methods.

1)\emph{Triple-duplicate (TD):} When the sender receives four ACKs with the same sequence number, then $W_{i+1} \leftarrow \frac{1}{2} W_i$.

2)\emph{Time-out (TO):} If the sender does not hear from the receiver for a predefined time period, called the ``time-out'' period (which is $T_o$ rounds long), then the sender closes its transmission window, $W_{i+1} \leftarrow 1$. At this point, the sender updates its TO period to $2T_o$ rounds, and transmits one packet. For any subsequent TO events, the sender transmits the one packet within its window, and doubles its TO period until $64T_o$ is reached, after which the TO period is fixed to $64T_o$. Once the sender receives an ACK from the receiver, it resets its TO period to $T_o$ and increments its window according to the congestion avoidance mechanism. During time-out, the throughput of both TCP and TCP/NC is zero.

Finally, in practice, the TCP receiver sends a single cumulative ACK after receiving $\beta$ number of packets, where $\beta = 2$ typically. However, we assume that $\beta = 1$ for simplicity. Extending the analysis to $\beta \geq 1$ is straightforward.

There are several variants to the traditional TCP congestion control. For example, STCP \cite{stcp} modifies the congestion control mechanism for networks with high bandwidth-delay products.
Other variants 
include those with selective acknowledgment schemes \cite{sack}. It may be interesting to compare the performance of the TCP variants with that of TCP/NC. However, we focus on traditional TCP here.

\subsection{Maximum window size}
In general, TCP cannot increase its window size unboundedly; there is a maximum window size $W_{\max}$. The TCP sender uses a congestion avoidance mechanism to increment the window size until $W_{\max}$, at which the window size remains $W_{\max}$ until a TD or a TO event.

\subsection{Erasures}

We assume that there is are random erasures within in the network. We denote $p$ to be the probability that a packet is lost at any given time. We assume that packet losses are independent. We note that this erasure model is different from that of \cite{TCP_Kurose} where losses are correlated within a round -- i.e. bursty erasures. Correlated erasures model well bursty traffic and congestion in wireline networks. In our case, however, we are aiming to model wireless networks, thus we shall use random independent erasures.

We do not model congestion or correlated losses within this framework, but show by simulation that when there are correlated losses, both TCP and TCP/NC close their window; thus, TCP/NC is able to react to congestion.

\subsection{Performance metric}

We analyze the performance of TCP and TCP/NC in terms of two metrics: the average throughput $\thru{}$, and the expected window evolution $E[W]$, where $\thru{}$ represents the total average throughput while window evolution $E[W]$ reflects the perceived throughput at a given time.
We define $\n{[t_1, t_2]}$ to be the number of packets received by the receiver during the interval $[t_1,t_2]$. The total average throughput is defined as:
\begin{equation}\label{eq:thru-era}
\thru{} = \lim_{\Delta \rightarrow \infty} \frac{\n{[t, t+\Delta]}}{\Delta}.
\end{equation}
We denote $\thru{tcp}$ and $\thru{nc}$ to be the average throughput for TCP and TCP/NC, respectively.

\subsection{Intuition}\label{sec:td-intuition}

For traditional TCP, random erasures in the network can lead to triple-duplicate ACKs. For example, in Figure \ref{fig:td-tcp}, the sender transmits $W_i$ packets in round $i$; however, only $a_i$ of them arrive at the receiver. As a result, the receiver ACKs the $a_i$ packets and waits for packet $a_i + 1$. When the sender receives the ACKs, round $i+1$ starts. The sender updates its window ($W_{i+1} \leftarrow W_i + a_i/W_i$), and starts transmitting the new packets in the window. However, since the receiver is still waiting for packet $a_i+1$, any other packets cause the receiver to request for packet $a_i +1$. This results in a triple-duplicate ACKs event and the TCP sender closes its window, i.e. $W_{i+2} \leftarrow \frac{1}{2} W_{i+1} = \frac{1}{2} (W_{i} + a_i/W_i)$.

Notice that this window closing due to TD does not occur when using TCP/NC as illustrated in Figure \ref{fig:td-tcpnc}. With network coding, any linearly independent packet delivers new information. Thus, any subsequent packet (in Figure \ref{fig:td-tcpnc}, the first packet sent in round $i+1$) can be viewed as packet $a_i + 1$. As a result, the receiver is able to increment its ACK and the sender continues transmitting data. It follows that network coding masks the losses within the network from TCP, and prevents it from closing its window by misjudging link losses as congestion. {\bf\emph{Network coding translates random losses as longer RTT}}, thus slowing down the transmission rate to adjust for losses without closing down the window in a drastic fashion.

Note that network coding does not mask correlated (or bursty) losses due to congestion.
With enough correlated losses, network coding cannot correct for all the losses. 
As a result, the transmission rate will be adjusted according to standard TCP's congestion control mechanism when TCP/NC detects correlated losses. Therefore, network coding allows TCP to maintain a high throughput connection in a lossy environment; at the same time, allows TCP to react to congestion. Thus, network coding naturally distinguishes congestion from random losses for TCP.

\section{Throughput Analysis for TCP}\label{sec:td}

We consider the effect of losses for TCP. The throughput analysis for TCP is similar to that of \cite{TCP_Kurose}. However, the model has been modified from that of \cite{TCP_Kurose} to account for independent losses and allow a fair comparison with network coded TCP. TCP can experience a TD or a TO event from random losses.

We note that, despite independent packet erasures, a single packet loss may affect subsequent packet reception. This is due to the fact that TCP requires in-order reception. A single packet loss within a transmission window forces all subsequent packets in the window to be out of order. Thus, they are discarded by the TCP receiver. As a result, standard TCP's throughput behavior with independent losses is similar to that of \cite{TCP_Kurose}, where losses are correlated within one round.


\begin{figure}[tbp]
\begin{center}
\includegraphics[width=0.40\textwidth]{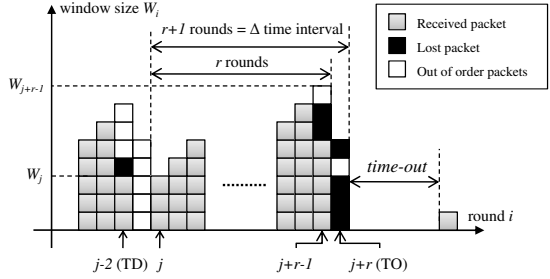}
\end{center}\vspace*{-.2cm}\caption{TCP's window size with a TD event and a TO event. In round $j-2$, losses occur resulting in triple-duplicate ACKs. On the other hand, in round $j+r-1$, losses occur; however, in the following round $j+r$ losses occur such that the TCP sender only receives two-duplicate ACKs. As a result, TCP experiences time-out.}\label{fig:td-rounds}\vspace*{-.2cm}
\end{figure}

\subsection{Triple-duplicate for TCP}\label{sec:td-tcp}

We consider the expected throughput between consecutive TD events, as shown in Figure \ref{fig:td-rounds}. Assume that the TD events occurred at time $t_1$ and $t_2 = t_1 + \Delta$, $\Delta > 0$. Assume that round $j$ begins immediately after time $t_1$, and that packet loss occurs in the $r$-th round, i.e. round $j+r-1$.

First, we calculate $E[\n{[t_1, t_2]}]$. Note that during the interval $[t_1, t_2]$, there are no packet losses. Given that the probability of a packet loss is $p$, the expected number of consecutive packets that are successfully sent from sender to receiver is
\begin{equation}\label{eq:td-packet}
E\left[\n{[t_1, t_2]}\right] = \left(\sum_{k=1}^\infty k (1-p)^{k-1} p \right)- 1= \frac{1-p}{p}.
\end{equation}

The packets (in white in Figure \ref{fig:td-rounds}) sent after the lost packets (in black in Figure \ref{fig:td-rounds}) are out of order, and will not be accepted by the standard TCP receiver. Thus, Equation (\ref{eq:td-packet}) does not take into account the packets sent in round $j-1$ or $j+r$.

We calculate the expected time period between two TD events, $E[\Delta]$. As in Figure \ref{fig:td-rounds}, after the packet losses in round $j$, there is an additional round for the loss feedback from the receiver to reach the sender. Therefore, there are $r+1$ rounds within the time interval $[t_1, t_2]$, and $\Delta = RTT ( r + 1)$. Thus,
\begin{equation}\label{eq:td-time}
E[\Delta] = RTT (E[r]+1).
\end{equation}
To derive $E[r]$, note that $W_{j+r-1} = W_{j}+ r-1$ and
\begin{equation}\label{eq:halfwindow}
W_j = \frac{1}{2} W_{j-1} = \frac{1}{2} \left(W_{j-2} + \frac{a_{j-2}}{W_{j-2}}\right).
\end{equation}
Equation (\ref{eq:halfwindow}) is due to TCP's congestion control. TCP interprets the losses in round $j-2$ as congestion, and as a result halves its window. Assuming that, in the long run, $E[W_{j+r-1}] = E[W_{j-2}]$ and that $a_{j-2}$ is uniformly distributed between $[0, W_{j-2}]$,
\begin{equation}\label{eq:td-window1}
E[W_{j+r-1}] = 2 \left(E[r] -\frac{3}{4}\right) \text{  and  } E[W_j] = E[r] -\frac{1}{2}.
\end{equation}
During these $r$ rounds, we expect to successfully transmit $\frac{1-p}{p}$ packets as noted in Equation (\ref{eq:td-packet}). This results in:
\begin{align}
\frac{1-p}{p}& 
= \left(\sum_{k=0}^{r-2} W_{j+k}\right) + a_{j+r-1}\label{eq:td-2}\\
& = (r-1)W_j + \frac{(r-1)(r-2)}{2} + a_{j+r-1}. \label{eq:td-1}
\end{align}
Taking the expectation of Equation (\ref{eq:td-1}) and using Equation (\ref{eq:td-window1}),
\begin{equation}\label{eq:quad}
\frac{1-p}{p} = \frac{3}{2}(E[r]-1)^2 + E[a_{j+r-1}].
\end{equation}
Note that $a_{j+r-1}$ is assumed to be uniformly distributed across $[0, W_{j+r-1}]$. Thus, $E[a_{j+r-1}] = E[W_{j+r-1}]/2 = E[r]-\frac{3}{4}$ by Equation (\ref{eq:td-window1}). Solving Equation (\ref{eq:quad}) for $E[r]$, we find:
\begin{equation}\label{eq:td-tcp-r}
E[r] = \frac{2}{3} + \sqrt{-\frac{1}{18} + \frac{2}{3}\frac{1-p}{p}}.
\end{equation}
The steady state
$E[W]$ is the average window size over two consecutive TD events. This provides an expression of steady state average window size for TCP (using Equations (\ref{eq:td-window1})):
\begin{align}
E[W] &= \frac{E[W_j] + E[W_{j+r-1}]}{2}=\frac{3}{2}E[r] -1. \label{eq:tcp-w}
\end{align}
The average throughput can be expressed as
\begin{equation}\label{eq:td-tcp-thru}
\thru{tcp}' = \frac{E[\n{[t_1, t_2]}]}{E[\Delta]} = \frac{1-p}{p}\frac{1}{RTT(E[r]+1)}.
\end{equation}
For small $p$, $\thru{tcp}' \approx \frac{1}{RTT}\sqrt{\frac{3}{2p}} + o(\frac{1}{\sqrt{p}})$; for large $p$, $\thru{tcp}' \approx \frac{1}{RTT}\frac{1-p}{p}$. If we only consider TD events, the long-term steady state throughput is equal to that in Equation (\ref{eq:td-tcp-thru}).

The analysis above assumes that the window size can grow unboundedly; however, this is not the case. To take maximum window size $W_{\max}$ into account, we make a following approximation:
\begin{equation}\label{eq:td-tcp}
\thru{tcp} = \min\left(\frac{W_{\max}}{RTT}, \thru{tcp}'\right).
\end{equation}
For small $p$, this result coincide with the results in \cite{TCP_Kurose}.

\subsection{Time-out for TCP}

If there are enough losses within two consecutive rounds, TCP may experience a TO event,
 as shown in Figure \ref{fig:td-rounds}.
Thus, $\mathbf{P}(\text{TO}|W)$, the probability of a TO event given a window size of $W$, is given by
\begin{equation}\label{eq:probTO}
\small
\mathbf{P}(\text{TO}|W) =
\begin{cases}
1 &\text{if $W <3$;}\\
\sum_{i=0}^2 \binom{W}{i} p^{W-i}(1-p)^i &\text{if $W \geq 3$}.
\end{cases}
\end{equation}
Note that when the window is small ($W <3$), then losses result in TO events. For example, assume $W=2$ with packets $\mathbf{p_1}$ and $\mathbf{p_2}$ in its window. Assume that $\mathbf{p_2}$ is lost. Then, the TCP sender may send another packet $\mathbf{p_3}$ in the subsequent round since the acknowledgment for $\mathbf{p_1}$ allows it to transmit a new packet. However, this would generate a single duplicate ACK with no further packets in the pipeline, and TCP sender waits for ACKs until it times out.

\begin{figure*}[tbp]
\small
\begin{align}
E[\text{duration of TO period}] &= (1-p) \left[T_o p  + 3T_o p^2 + 7 T_o p^3  + 15 T_o p^4 + 31 T_o p^5  + \sum_{i=0}^{\infty} (63 + i\cdot 64)T_o p^{6 + i}\right]\\
&= (1-p) \left[T_o p  + 3T_o p^2 + 7 T_o p^3  + 15 T_o p^4 + 31 T_o p^5  + 63T_o\frac{p^6}{1-p} + 64T_o\frac{p^7}{(1-p)^2} \right] \label{eq:down}
\end{align}\vspace*{-.3cm}
\end{figure*}


\begin{figure*}[tbp]
\small
\begin{equation}\label{eq:tcp}
\thru{tcp} = \min\left(\frac{W_{\max}}{RTT}, \frac{1-p}{p}\frac{1}{RTT\left(\frac{5}{3} + \sqrt{-\frac{1}{18} + \frac{2}{3}\frac{1-p}{p}} + \mathbf{P}(\text{TO}|E[W])E[\text{duration of TO period}]\right)} \right)
\end{equation}\vspace*{-.3cm}
\end{figure*}

We approximate $W$ in above Equation (\ref{eq:probTO}) with the expected window size $E[W]$ from Equation (\ref{eq:tcp-w}). The length of the TO event depends on the duration of the loss events. Thus, the expected duration of TO period (in RTTs) is given in Equation (\ref{eq:down}). Finally, by combining the results in Equations (\ref{eq:td-tcp}), (\ref{eq:probTO}), and (\ref{eq:down}), we get an expression for the average throughput of TCP as shown in Equation (\ref{eq:tcp}).

\section{Throughput Analysis for TCP/NC}\label{sec:tcpnc}

\begin{figure}[tbp]
\begin{center}
\includegraphics[width=0.40\textwidth]{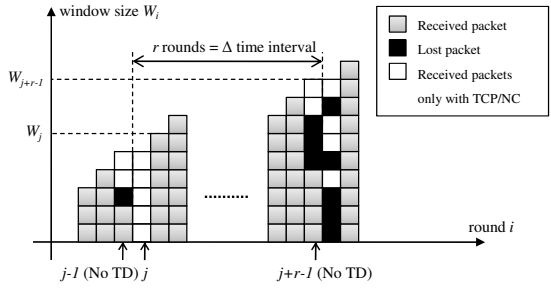}
\end{center}\vspace*{-.2cm}\caption{TCP/NC's window size with erasures that would lead to a triple-duplicate ACKs event when using standard TCP. Note that unlike TCP, the window size is non-decreasing.}\label{fig:td-tcpncrounds}
\end{figure}

\begin{figure}[tbp]
\begin{center}
\includegraphics[width=0.35\textwidth]{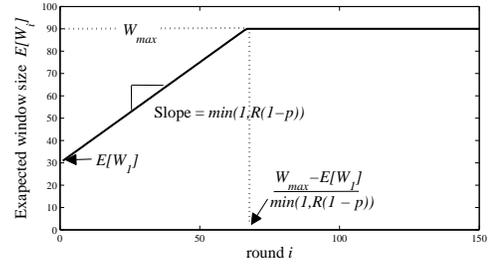}
\end{center}\vspace*{-.2cm}\caption{Expected window size for TCP/NC where $W_{\max} = 90$, $E[W_1] = 30$. We usually assume $E[W_1] = 1$; here we use $E[W_1] = 30$ to exemplify the effect of $E[W_1]$.}\label{fig:period}
\end{figure}

\begin{figure}[t]
\begin{center}
\includegraphics[width=0.5\textwidth]{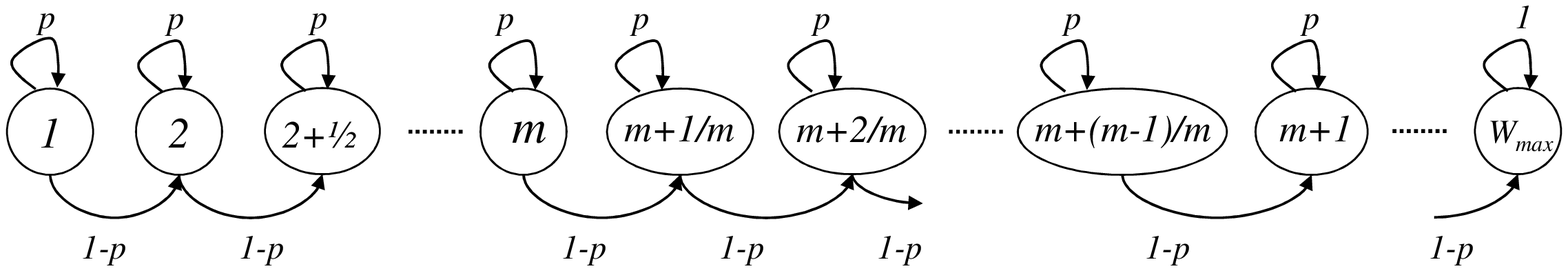}
\end{center}\vspace*{-.2cm}\caption{Markov chain for the TCP/NC's window evolution.}\vspace*{-.2cm}\label{fig:markov}
\end{figure}

\begin{figure}[tbp]
\[ P = \scriptsize
\left(
\begin{array}{cccccccccc}
\cellcolor[gray]{.8} p &\cellcolor[gray]{.8} 1-p &\cellcolor[gray]{.8}  0 & \cellcolor[gray]{.8} 0 &\cellcolor[gray]{.8}  0 &\cellcolor[gray]{.8} \cdots & \cellcolor[gray]{.8} 0 & 0  \\
\cellcolor[gray]{.8} 0 & \cellcolor[gray]{.8} p & \cellcolor[gray]{.8} 1-p &\cellcolor[gray]{.8}  0 &\cellcolor[gray]{.8}  0 &\cellcolor[gray]{.8}  \dotsb &\cellcolor[gray]{.8}  0 & 0  \\
\cellcolor[gray]{.8} 0 & \cellcolor[gray]{.8} 0 &\cellcolor[gray]{.8}  p & \cellcolor[gray]{.8} 1-p & \cellcolor[gray]{.8} 0&\cellcolor[gray]{.8}  \dotsb &\cellcolor[gray]{.8}  0 & 0  \\
\cellcolor[gray]{.8} \vdots & \cellcolor[gray]{.8}  &  \cellcolor[gray]{.8} & \cellcolor[gray]{.8} \ddots & \cellcolor[gray]{.8} \ddots&  \cellcolor[gray]{.8} \cdots &\cellcolor[gray]{.8} & \vdots \\
\cellcolor[gray]{.8} 0 &\cellcolor[gray]{.8}  0 & \cellcolor[gray]{.8} 0 & \cellcolor[gray]{.8} 0 & \cellcolor[gray]{.8} 0 &\cellcolor[gray]{.8} 0& \cellcolor[gray]{.8}  p & 1-p\\
0 & 0 & 0 & 0 & 0 & 0&0 & 1\\
\end{array}
\right)
\]\caption{The \emph{transition matrix} $P$ for the Markov chain in Figure \ref{fig:markov}. The shaded part of the matrix is denoted $Q$. Matrix $N = (I-Q)^{-1}$ is the \emph{fundamental matrix} of the Markov chain, and can be used to compute the expected rounds until the absorption state.}\label{fig:transition}
\end{figure}

We consider the expected throughput for TCP/NC. Note that erasure patterns that result in TD and/or TO events under TCP may not yield the same result under TCP/NC, as illustrated in Section \ref{sec:td-intuition}. We emphasize again that this is due to the fact that any linearly independent packet conveys a new degree of freedom to the receiver. Figure \ref{fig:td-tcpncrounds} illustrates this effect -- packets (in white) sent after the lost packets (in black) are acknowledged by the receivers, thus allowing TCP/NC to advance its window. This implies that TCP/NC does not experience window closing owing to random losses often.


\subsection{TCP/NC Window Evolution}\label{sec:tcpnc-window}

From Figure \ref{fig:td-tcpncrounds}, we observe that TCP/NC is able to maintain its window size despite experiencing losses. This is partially because TCP/NC is able to receive packets that would be considered out of order by TCP. As a result, TCP/NC's window evolves differently from that of TCP, and can be characterized by a simple recursive relationship as
\begin{equation*}\label{eq:tcpnc-w}
E[W_i] = E[W_{i-1}] + \frac{E[a_{i-1}]}{E[W_{i-1}]} = E[W_{i-1}] + \min\{1, R(1-p)\}.
\end{equation*}
The recursive relationship captures the fact that every packet that is linearly independent of previously received packets is considered to be \emph{innovative} and is therefore acknowledged. Consequently, any arrival at the receiver is acknowledged with high probability; thus, we expect $E[a_{i-1}]$ packets to be acknowledged and the window to be incremented by $ \frac{E[a_{i-1}]}{E[W_{i-1}]}$. Note that $E[a_{i-1}] = (1-p)\cdot R\cdot  E[W_{i-1}]$ since the encoder transmits on average $R$ linear combinations for every packet transmitted by the TCP sender.

Once we take
 $W_{\max}$ into account, we have the following expression for TCP/NC's expected window size:
\begin{equation}\label{eq:tcpnc-w2}
E[W_i] = \min(W_{\max}, E[W_1] + i\min\{1, R(1-p)\}),
\end{equation}
where $i$ is the round number. $E[W_1]$ is the initial window size, and we set $E[W_1] = 1$. Figure \ref{fig:period} shows an example of the evolution of the TCP/NC window using Equation (\ref{eq:tcpnc-w2}).

\subsubsection{Markov Chain Model}\label{sec:markov-chain-model}

The above analysis describes the expected behavior of TCP/NC's window size. We can also describe the window size behavior using a Markov chain as shown in Figure \ref{fig:markov}. The states of this Markov chain represent the instantaneous window size (not specific to a round). A transition occurs whenever a packet is transmitted. We denote $S(W)$ to be the state representing the window size of $W$. Assume that we are at state $S(W)$. If a transmitted packet is received by the TCP/NC receiver and acknowledged, the window is incremented by $\frac{1}{W}$; thus, we end up in state $S(W + \frac{1}{W})$. This occurs with probability $(1-p)$. On the other hand, if the packet is lost, then we stay at $S(W)$. This occurs with probability $p$. Thus, the Markov chain states represent the window size, and the transitions correspond to packet transmissions.

Note that $S(W_{\max})$ is an absorbing state of the Markov chain. As noted in Section \ref{sec:td-intuition}, TCP/NC does not often experience a window shutdown, which implies that there are correlated or heavy losses. Thus, TCP/NC's window size is non-decreasing, as shown in Figure \ref{fig:markov}. Therefore, given enough time, TCP/NC reaches state $S(W_{\max})$ with probability equal to 1. We analyze the expected number of packet transmissions needed for absorption.

The \emph{transition matrix} $P$ and the \emph{fundamental matrix} $N = (I-Q)^{-1}$ of the Markov chain is given in Figure \ref{fig:transition}. The entry $N(S_1, S_2)$ represents the expected number of visits to state $S_2$ before absorption -- i.e. we reach state $S(W_{\max})$ -- when we start from state $S_1$. Our objective is to find the expected number of packets transmitted to reach $S(W_{\max})$ starting from state $S(E[W_1])$ where $E[W_1]=1$. The partial sum of the first row entries of $N$ gives the expected number of packets transmitted until we reach the window size $W$. The expression for the first row of $N$ can be derived using cofactors: $N(1,:) = \left[\frac{1}{1-p}, \frac{1}{1-p}, \cdots, \frac{1}{1-p}\right]$.
The expected number of packet transmissions $T_p(W)$ to reach a window size of $W \in [1, W_{\max}]$ is:
\begin{align}
\nonumber T_p(W) &= \sum_{m = S(1)}^{S(W)} N(1, m) = \sum_{m=S(1)}^{S(W)} \frac{1}{1-p}= \frac{1}{1-p}\sum_{m=S(1)}^{S(W)} 1 \\
&=\frac{W(W-1)}{2(1-p)}.\label{eq:R}
\end{align}

$T_p(W)$ is the number of packets we expect to transmit given the erasure probability $p$. If we set $p = 0$, then $T_{0}(W) = \frac{W(W-1)}{2}$. Therefore, $\frac{W(W-1)}{2}$ is the minimal number of transmission needed to achieve $W$ (since this assumes no packets are lost). Note that $\frac{T_p(W)}{T_{0}(W)} = \frac{1}{1-p}$ represents a lower bound on cost when losses are introduced -- i.e. to combat random erasures, the sender on average has to send at least $\frac{1}{1-p}$ packets for each packet it wishes to send. This is exactly the definition of redundancy factor $R$. This analysis indicates that we should set $R \geq \frac{ T_p(W)}{T_{0}(W)}$. Furthermore, $T_{0}(W)$ is equal to the area under the curve for rounds $i \in [0, \frac{W-E[W_1]}{\min\{1, R\cdot(1-p)\}}]$ in Figure \ref{fig:period} if we set $R \geq \frac{1}{1-p}$. A more detailed discussion on the effect of $R$ is in Section \ref{sec:redundancy}.

\subsection{TCP/NC Analysis per Round}\label{sec:e2e-round}

Using the results in Section \ref{sec:tcpnc-window}, we derive an expression for the throughput.
 The throughput of round $i$, $\thru{i}$, is directly proportional to the window size $E[W_i]$, i.e.
\begin{equation}\label{eq:nc-throughput-round}
\thru{i}  =\frac{E[W_i]}{SRTT}  \min\{1, R(1-p)\}\text{ packets per second,}
\end{equation}
where 
 $SRTT$ is the round trip time estimate. The $RTT$ and its estimate $SRTT$ play an important role in TCP/NC. We shall formally define and discuss the effect of $R$ and $SRTT$ below.

We note that $\thru{i} \propto (1-p)\cdot R\cdot E[W_i]$. At any given round $i$, TCP/NC sender transmits $R\cdot E[W_i]$ coded packets,
 and we expect $pR\cdot E[W_i]$ packets to be lost. Thus, the TCP/NC receiver only receives $(1-p)\cdot R\cdot E[W_i]$ degrees of freedom.

\subsubsection{Redundancy Factor $R$}\label{sec:redundancy}
The redundancy factor $R \geq 1$ is the ratio between the average rate at which linear combinations are sent to the receiver and the rate at which TCP's window progresses. For example, if the TCP sender has 10 packets in its window, then the encoder transmits $10R$ linear combinations. If $R$ is large enough, the receiver will receive at least $10$ linear combinations to decode the original 10 packets.
 This redundancy is necessary to (a) compensate for the losses within the network, and (b) match TCP's sending rate to the rate at which data is actually received at the receiver. References \cite{tcpnc,tcpnc2} introduce the redundancy factor with TCP/NC, and show that $R \geq \frac{1}{1-p}$ is necessary. This coincides with our analysis in Section \ref{sec:markov-chain-model}.


The redundancy factor $R$ should be chosen with some care.
If $R < \frac{1}{1-p}$ causes significant performance degradation, since network coding can no longer fully compensate for the losses which may lead to window closing for TCP/NC. To maximize throughput, an optimal value of $R \geq \frac{1}{1-p}$ should be chosen. However, setting $R \gg \frac{1}{1-p}$ may over-compensate for the losses within the network; thus, introducing more redundant packets than necessary. On the other hand, matching $R$ to exactly $\frac{1}{1-p}$ may not be desirable for two reasons: 1) The exact value of $\frac{1}{1-p}$ may not be available or difficult to obtain in real applications; 2) As $R \rightarrow \frac{1}{1-p}$, it becomes more likely that TCP/NC is unable to \emph{fully} recover from losses in any given round. By \emph{fully} recover, we mean that TCP/NC decoder is able to acknowledge all packet transmitted in that round. As we shall show in Section \ref{sec:simulations}, TCP/NC can maintain a fairly high throughput with just partial acknowledgment (in each round, only a subset of the packets are acknowledged owing to losses). However, we still witness a degradation in throughput as $R$ decreases. Thus, we assume that $R\geq \frac{1}{1-p}$.

\subsubsection{Effective Round Trip Time $SRTT$}\label{sec:srtt}

$SRTT$ is the round trip time estimate that TCP maintains by sampling the behavior of packets sent over the connection. It is denoted $SRTT$ because it is often referred to as ``smoothed'' round trip time as it is obtained by averaging the time for a packet to be acknowledged after the packet has been sent. We note that, in Equation (\ref{eq:nc-throughput-round}), we use $SRTT$ instead of $RTT$ because $SRTT$ is the ``effective'' round trip time TCP/NC experiences.

In lossy networks, TCP/NC's $SRTT$ is often greater than $RTT$. This can be seen in Figure \ref{fig:example}. The first coded packet ($\mathbf{p_1 + p_2}$ $\mathbf{+ p_3}$) is received and acknowledged ($\mathbf{seen(p_1)}$). Thus, the sender is able to estimate the round trip time correctly; resulting in $SRTT = RTT$. However, the second packet ($\mathbf{p_1 +}$ $\mathbf{2p_2 + p_3}$) is lost. As a result, the third packet ($\mathbf{p_1 + 2p_2 +}$ $\mathbf{2 p_3}$) is used to acknowledge the second degree of freedom ($\mathbf{seen(p_2)}$). In our model, we assume for simplicity that the time needed to transmit a packet is much smaller than RTT; thus, despite the losses, our model would result in $SRTT \approx RTT$. However, in practice, depending on the size of the packets, the transmission time may not be negligible.



\subsection{TCP/NC Average Throughput}\label{sec:e2e-average}

Taking Equation (\ref{eq:nc-throughput-round}), we can average the throughput over $n$ rounds to obtain the average throughput for TCP/NC.
\begin{align}
\nonumber \thru{nc} &= \frac{1}{n}\sum_{i=1}^n \frac{E[W_i]}{SRTT}\min\{1, R(1-p)\}\\
&= \frac{1}{n\cdot SRTT} \cdot f(n),\label{eq:nc-final-longterm}
\end{align}
where
\begin{align*}
f(n) &=
\begin{cases} nE[W_1] + \frac{n(n+1)}{2} \text{\hspace*{.5cm} for $n \leq r^*$}\\
nW_{\max} - r^*(W_{\max} - E[W_1]) + \frac{r^*(r^*-1)}{2} \text{ for $n > r^*$}\\
\end{cases}\\
r^* &= W_{\max} - E[W_1].
\end{align*}
Note that as $n \rightarrow \infty$, the average throughput $\thru{nc} \rightarrow \frac{W_{\max}}{SRTT}$.

An important aspect of TCP is congestion control mechanism. This analysis may suggest that network coding no longer allows for TCP to react to congestion. We emphasize that the above analysis assumes that there are only random losses with probability $p$, and that there are no correlated losses. It is important to note that the erasure correcting power of network coding is limited by the redundancy factor $R$.
If there are enough losses (e.g., losses caused by congestion),
network coding cannot mask all the erasures from TCP. This may lead TCP/NC to experience a TD or TO event, depending on the variants of TCP used. In Section \ref{sec:congestioncontrol}, we present simulation results that show that TCP's congestion control mechanism still applies to TCP/NC when appropriate.




\section{Simulation Results for Throughput Analysis}\label{sec:simulations}

\begin{figure}[tbp]
\begin{center}\vspace*{.3cm}
\includegraphics[width=0.35\textwidth]{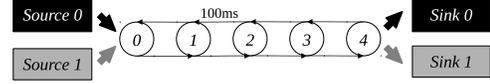}
\end{center}\caption{Network topology for the simulations.}\label{fig:setup}\vspace*{-.2cm}
\end{figure}

We use simulations to verify that our analysis captures the behavior of both TCP and TCP/NC. We use NS-2 (Network Simulator \cite{ns}) to simulate TCP and TCP/NC, where
we use the implementation of TCP/NC from \cite{tcpnc2}. Two FTP applications (ftp0, ftp1) wish to communicate from the source (src0, src1) to sink (sink0, sink1), respectively. There is no limit to the file size. The sources generate packets continuously until the end of the simulation. The two FTP applications use either TCP or TCP/NC. We denote TCP0, TCP1 to be the two FTP applications when using TCP; and we denote NC0, NC1 to be the two FTP applications when using TCP/NC.

\begin{figure*}[tbp]
\begin{center}
\subfloat[$p=0$]{\includegraphics[width=0.19\textwidth]{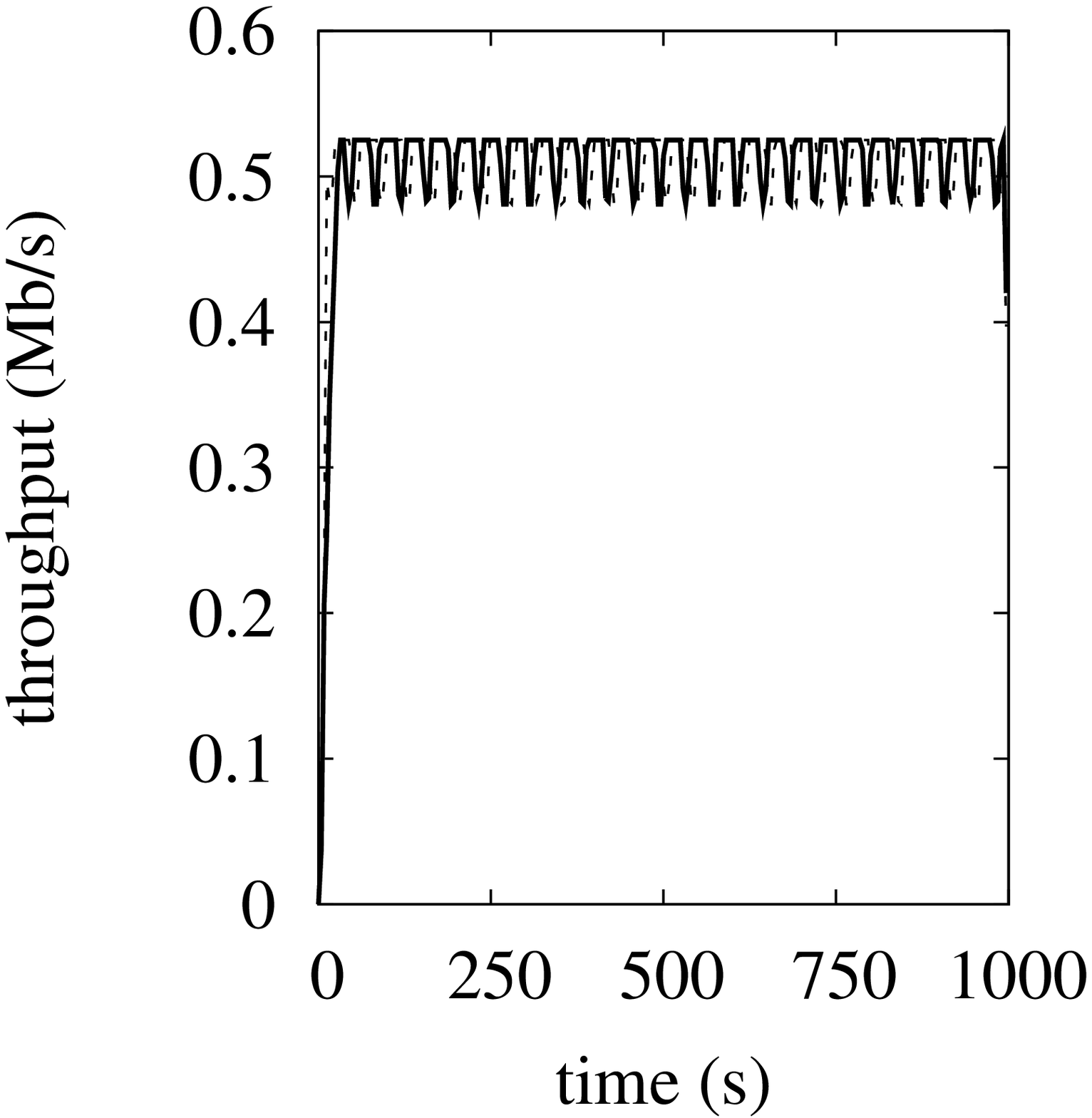}\label{fig:bw_0}}
\subfloat[$p=0.0199$]{\includegraphics[width=0.19\textwidth]{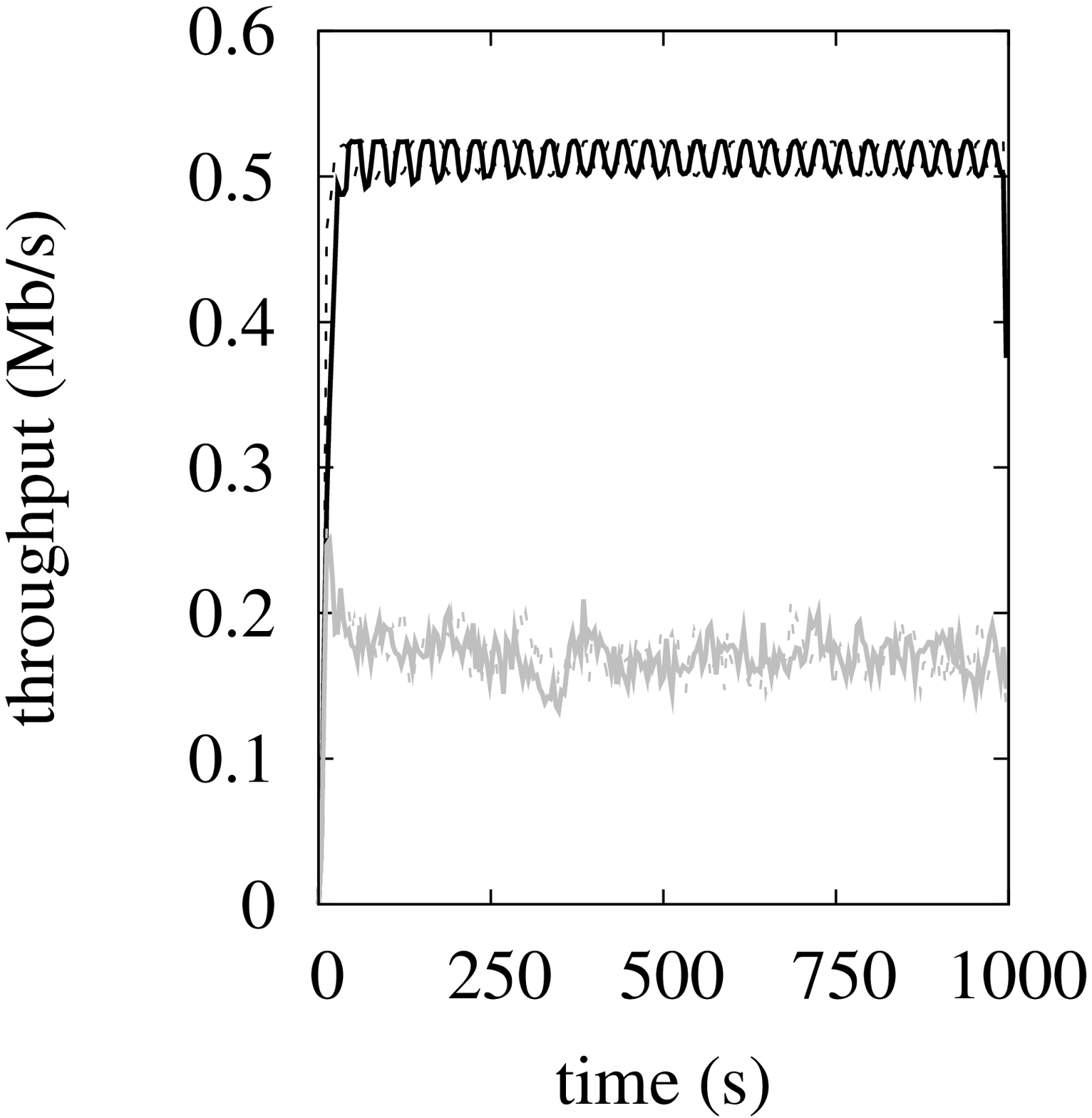}\label{fig:bw_5}}
\subfloat[$p=0.0587$]{\includegraphics[width=0.19\textwidth]{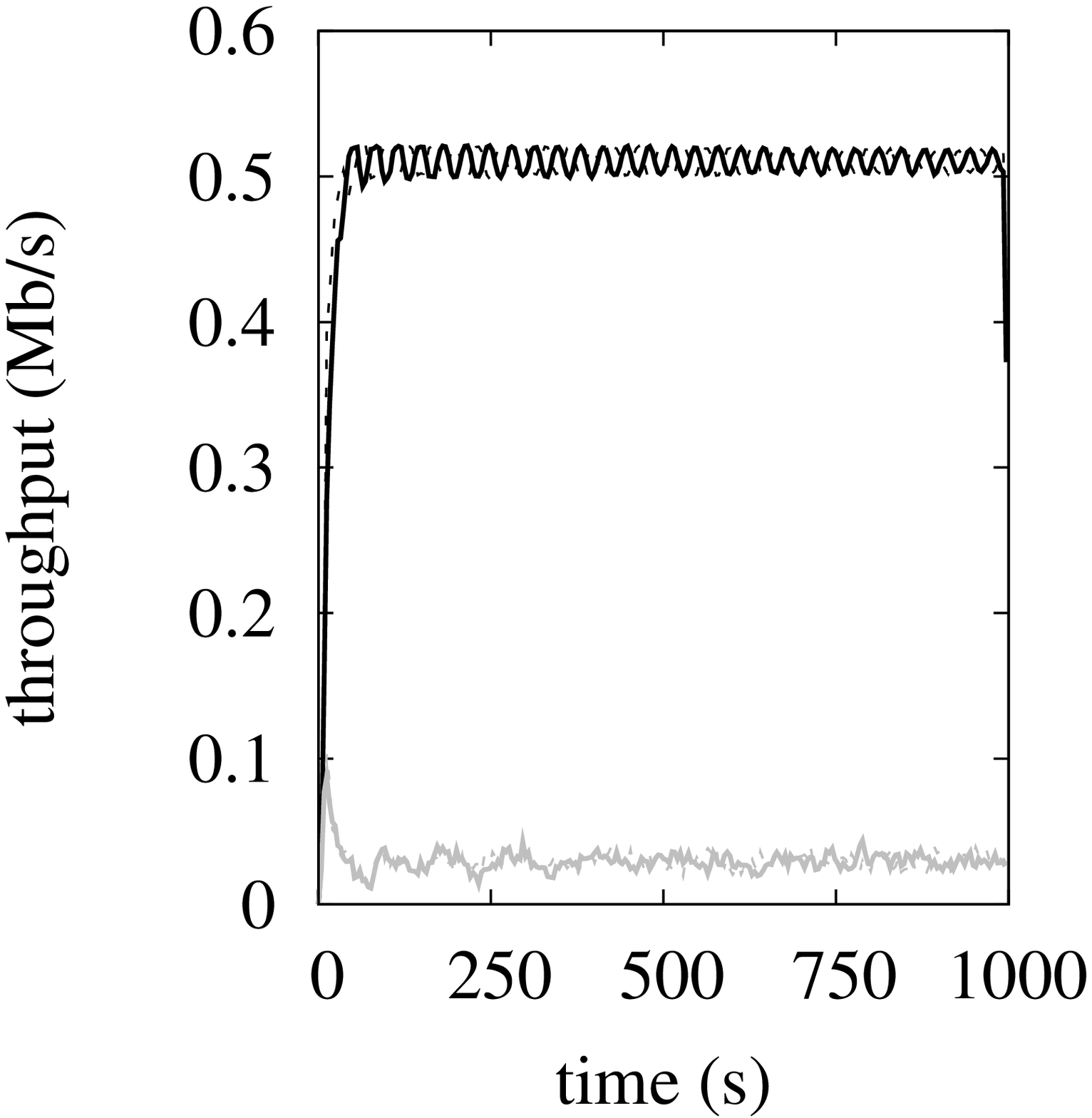}\label{fig:bw_15}}
\subfloat[$p=0.0963$]{\includegraphics[width=0.19\textwidth]{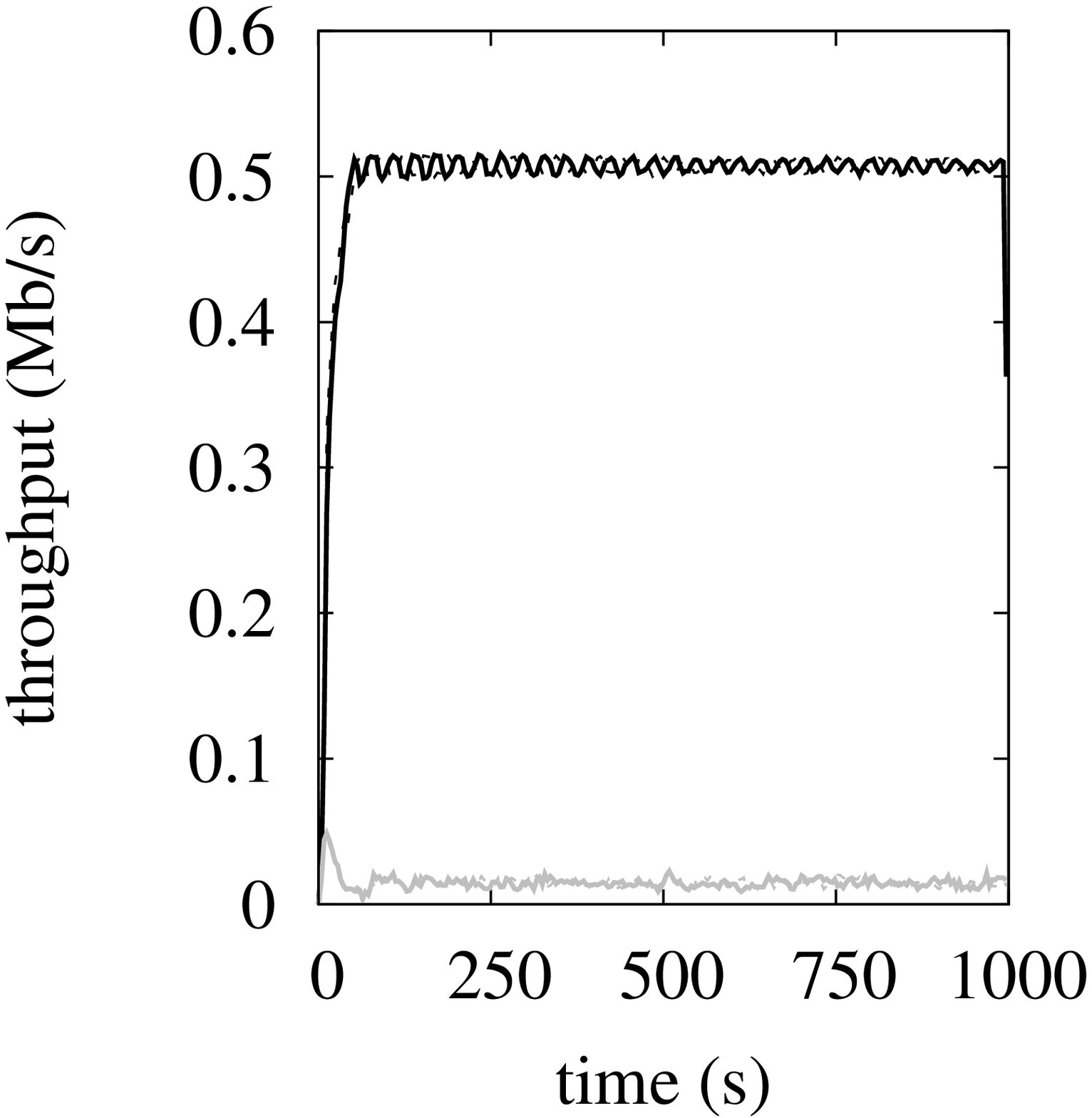}\label{fig:bw_25}}
\subfloat[$p=0.1855$]{\includegraphics[width=0.19\textwidth]{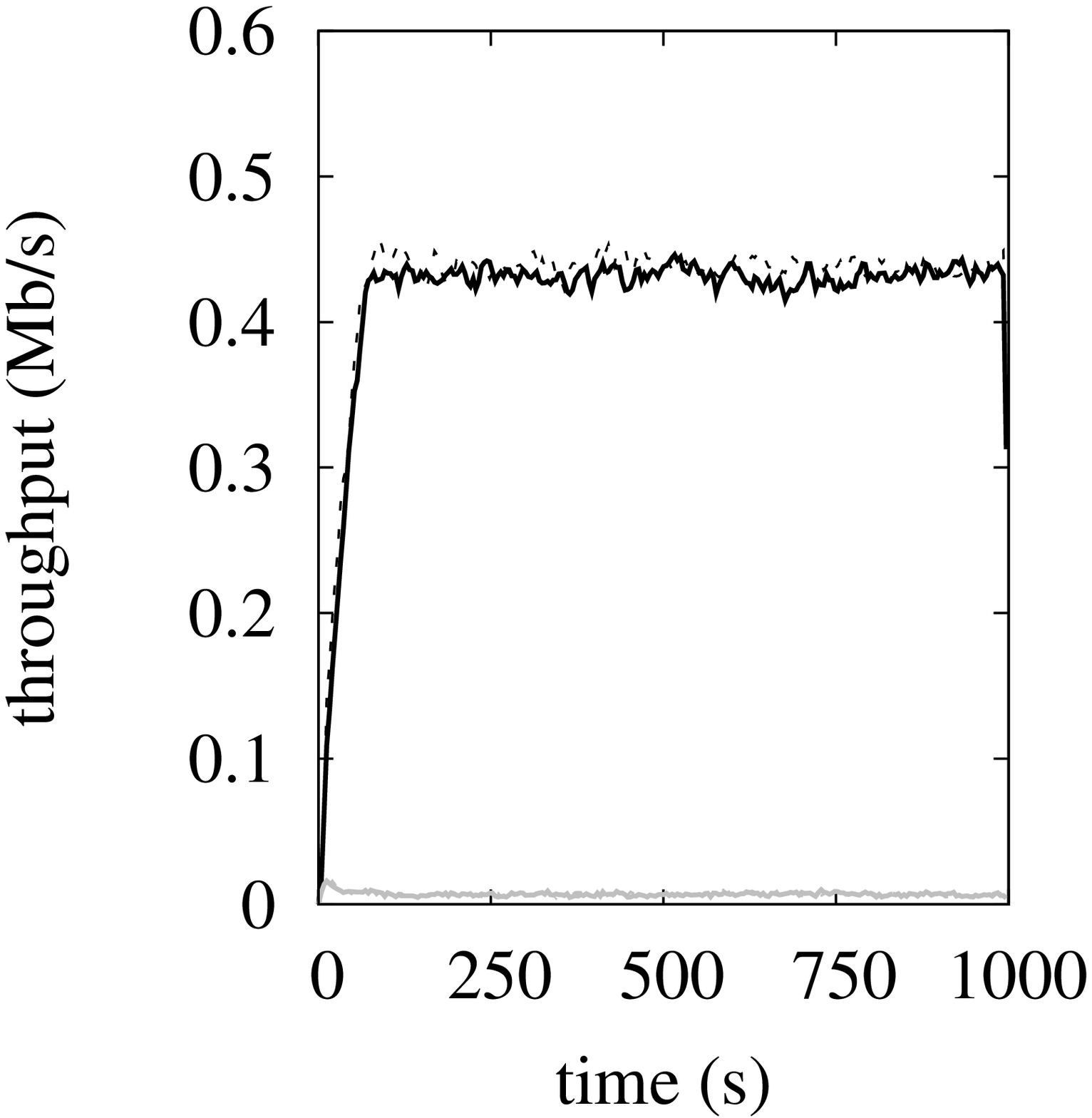}\label{fig:bw_50}}
\subfloat{\includegraphics[width=0.065\textwidth]{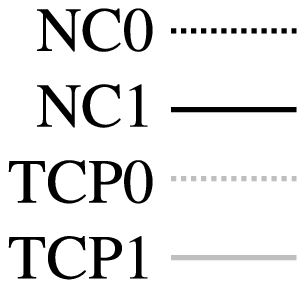}}
\end{center}\vspace*{-.2cm}\caption{Throughput of TCP/NC and TCP with varying link erasure probability $p$.}\label{fig:bw}\vspace*{-.2cm}
\end{figure*}

\begin{figure*}[tbp]
\begin{center}
\subfloat[$p=0$]{\includegraphics[width=0.19\textwidth]{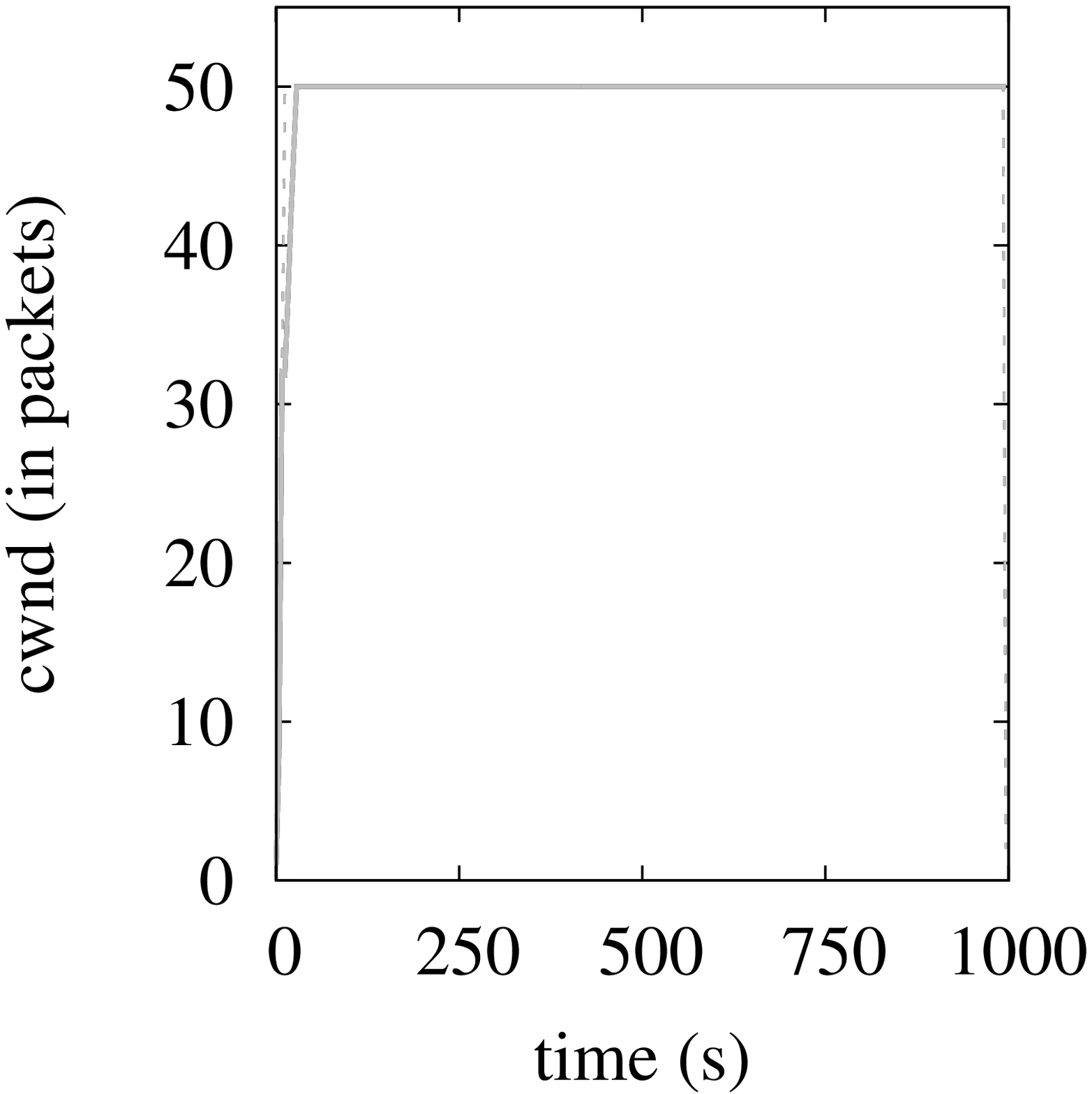}\label{fig:cw_0}}
\subfloat[$p=0.0199$]{\includegraphics[width=0.19\textwidth]{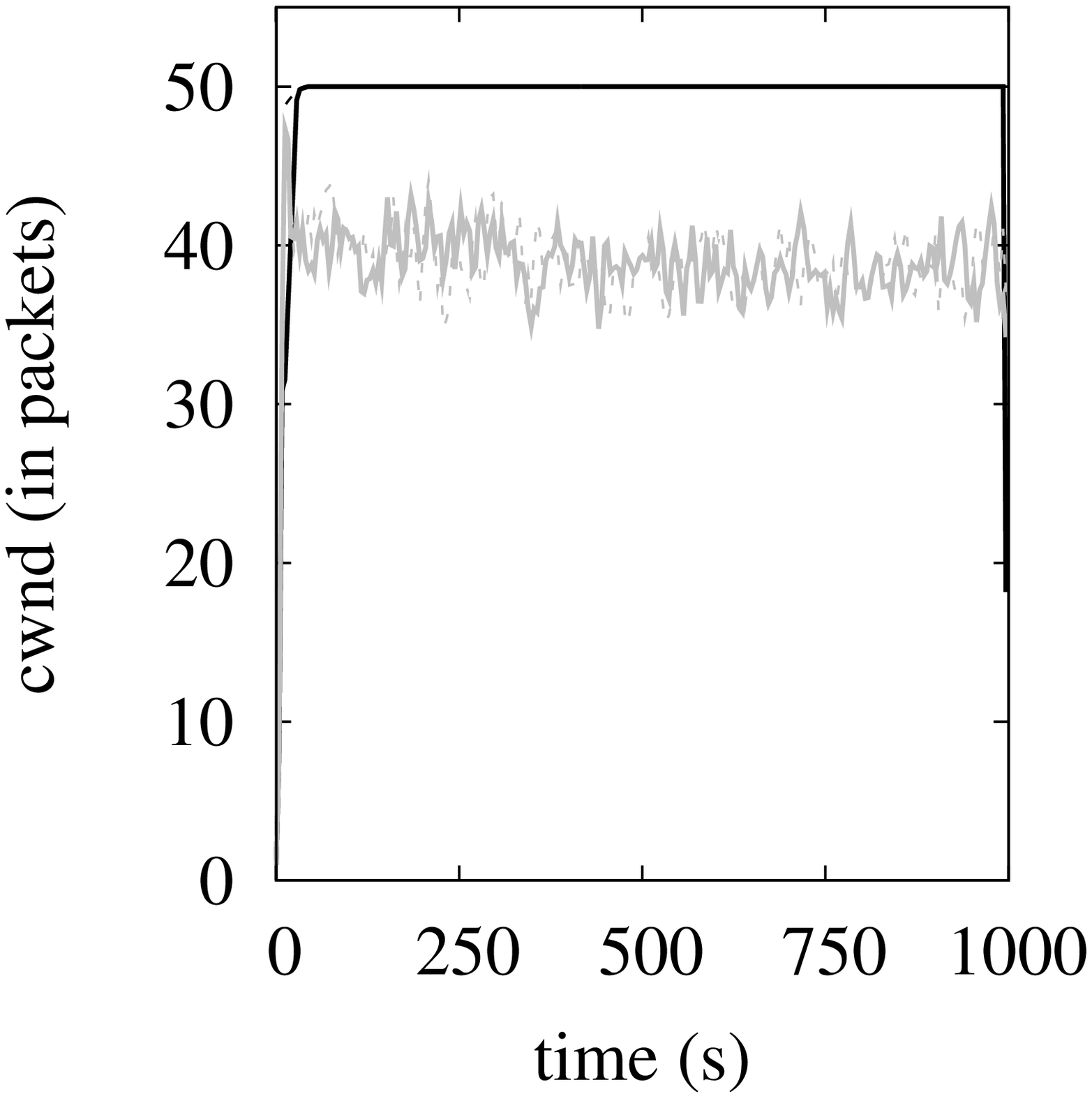}\label{fig:cw_5}}
\subfloat[$p=0.0587$]{\includegraphics[width=0.19\textwidth]{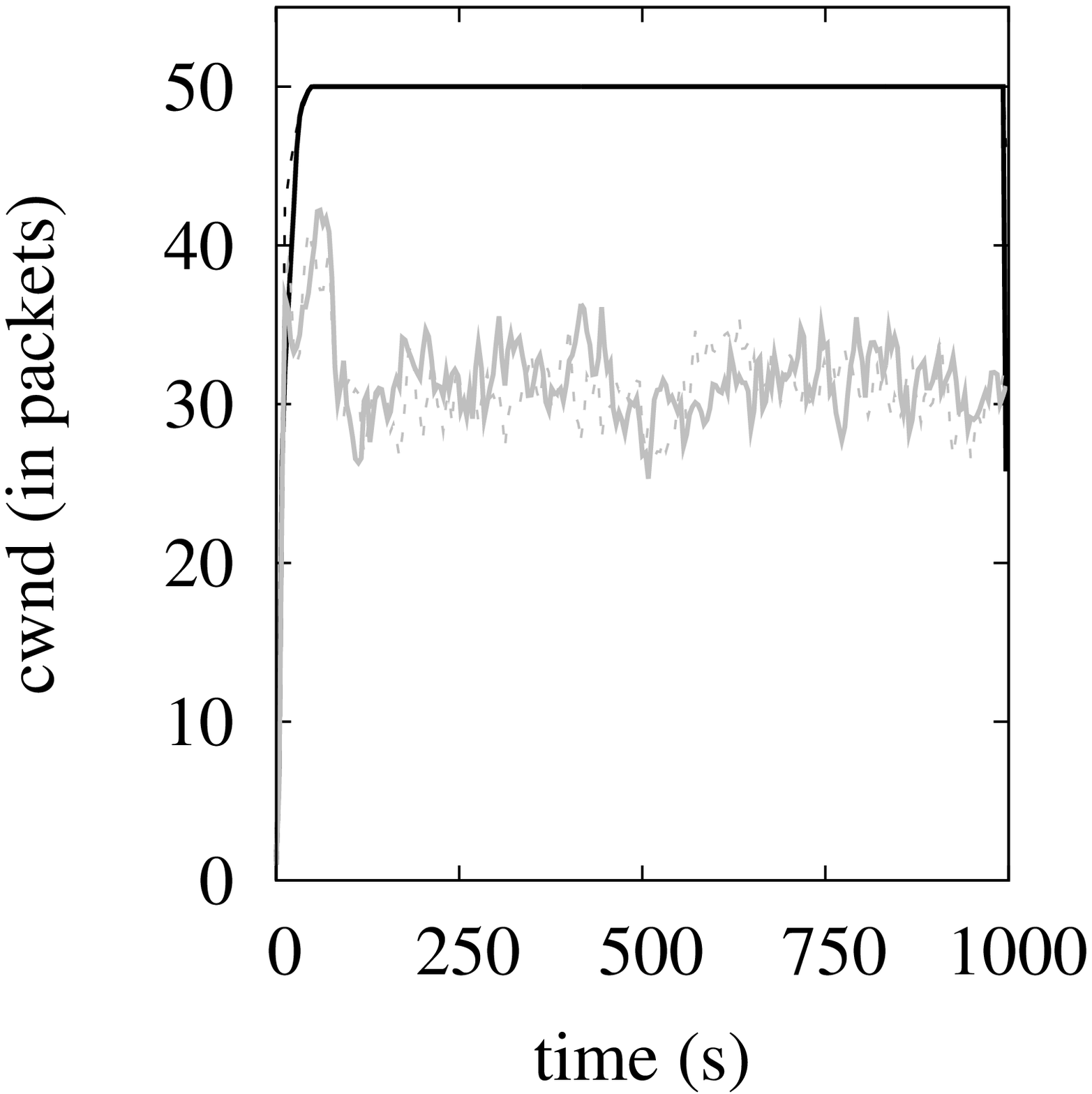}\label{fig:cw_15}}
\subfloat[$p=0.0963$]{\includegraphics[width=0.19\textwidth]{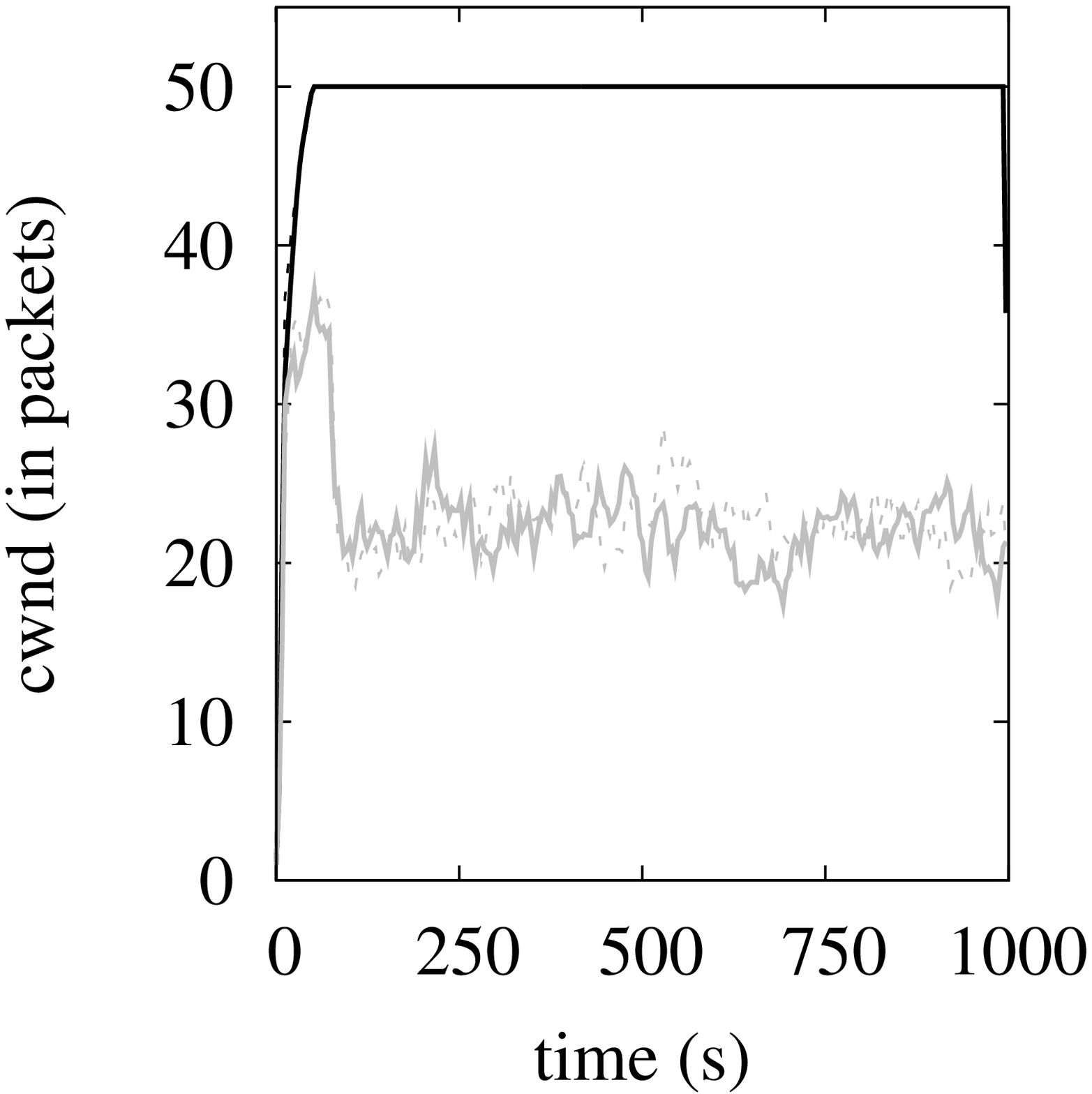}\label{fig:cw_25}}
\subfloat[$p=0.1855$]{\includegraphics[width=0.19\textwidth]{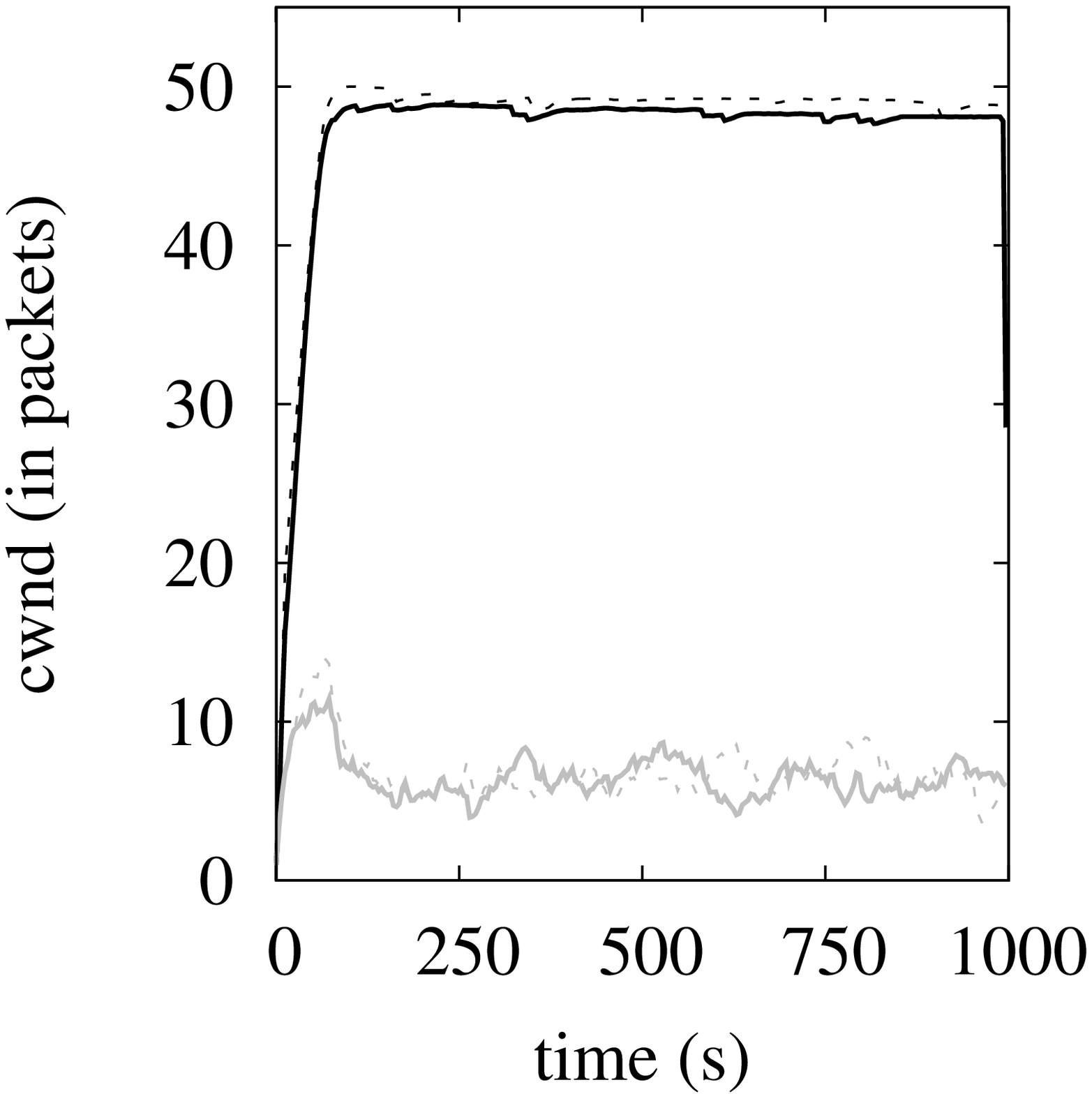}\label{fig:cw_50}}
\subfloat{\includegraphics[width=0.065\textwidth]{legend}}
\end{center}\vspace*{-.2cm}\caption{The congestion window size of TCP/NC and TCP with varying link erasure probability $p$.}\label{fig:cw}\vspace*{-.2cm}
\end{figure*}

\begin{figure*}[tbp]
\begin{center}
\subfloat[$p=0$]{\includegraphics[width=0.19\textwidth]{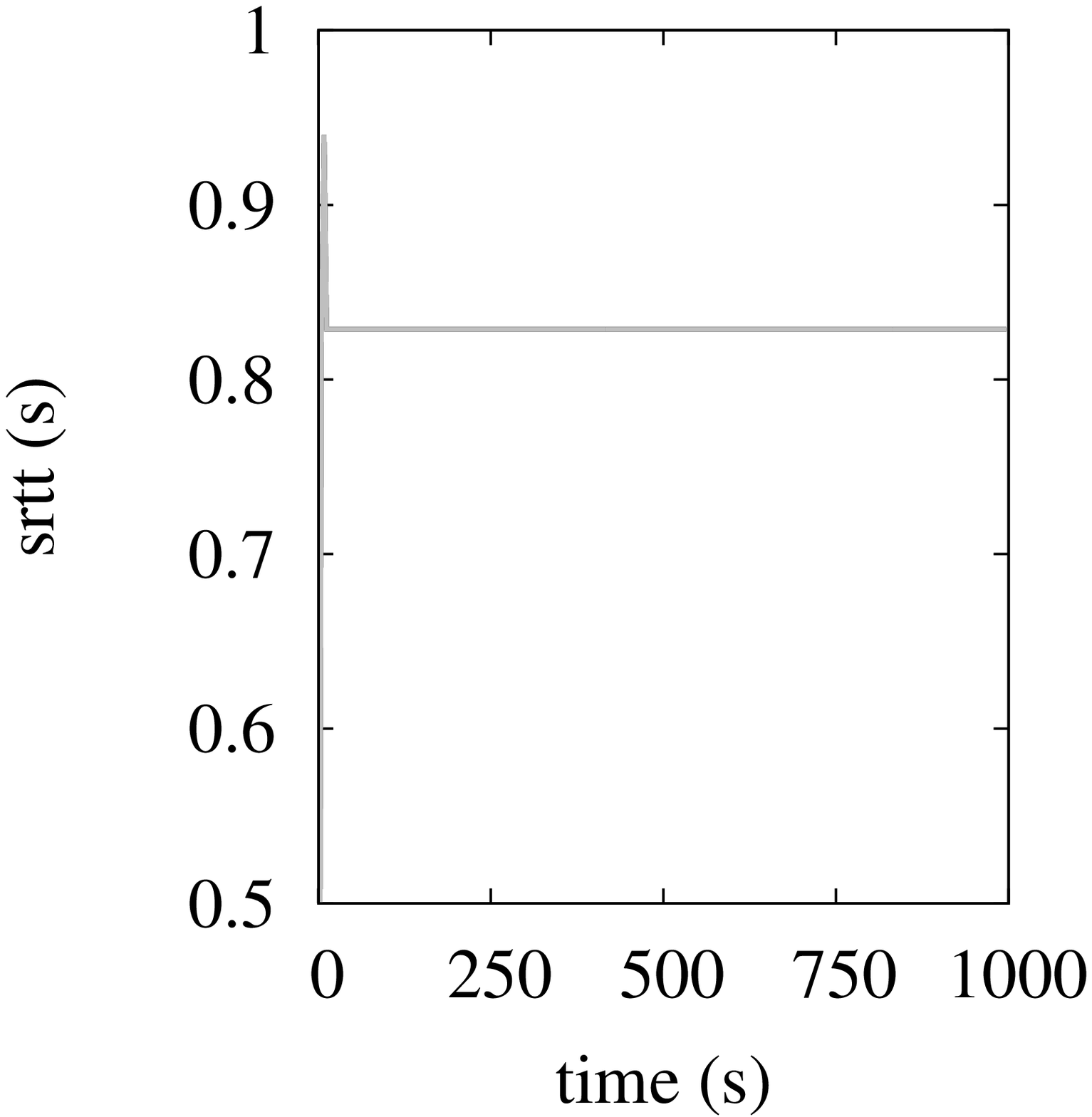}\label{fig:rtt_0}}
\subfloat[$p=0.0199$]{\includegraphics[width=0.19\textwidth]{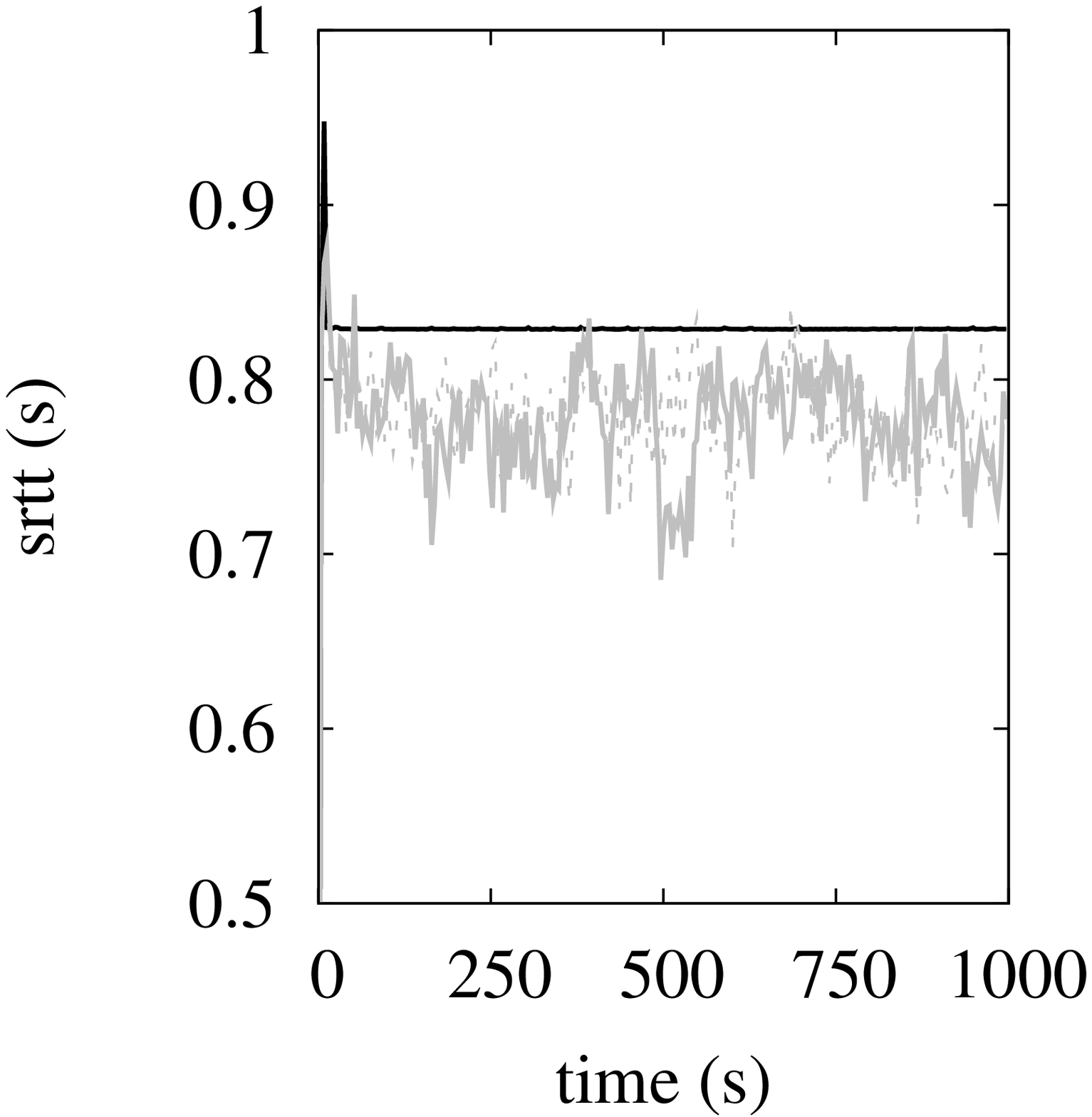}\label{fig:rtt_5}}
\subfloat[$p=0.0587$]{\includegraphics[width=0.19\textwidth]{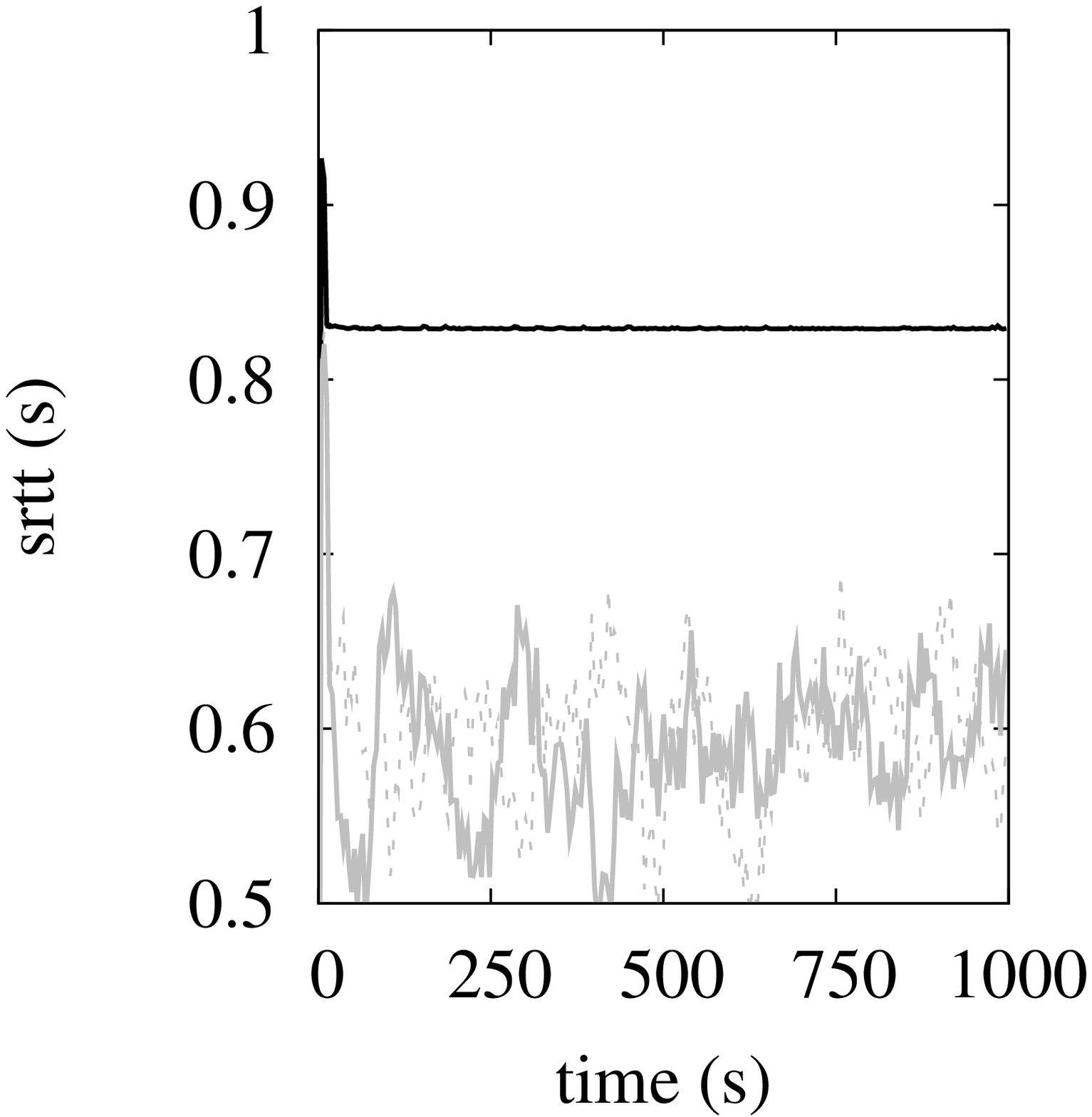}\label{fig:rtt_15}}
\subfloat[$p=0.0963$]{\includegraphics[width=0.19\textwidth]{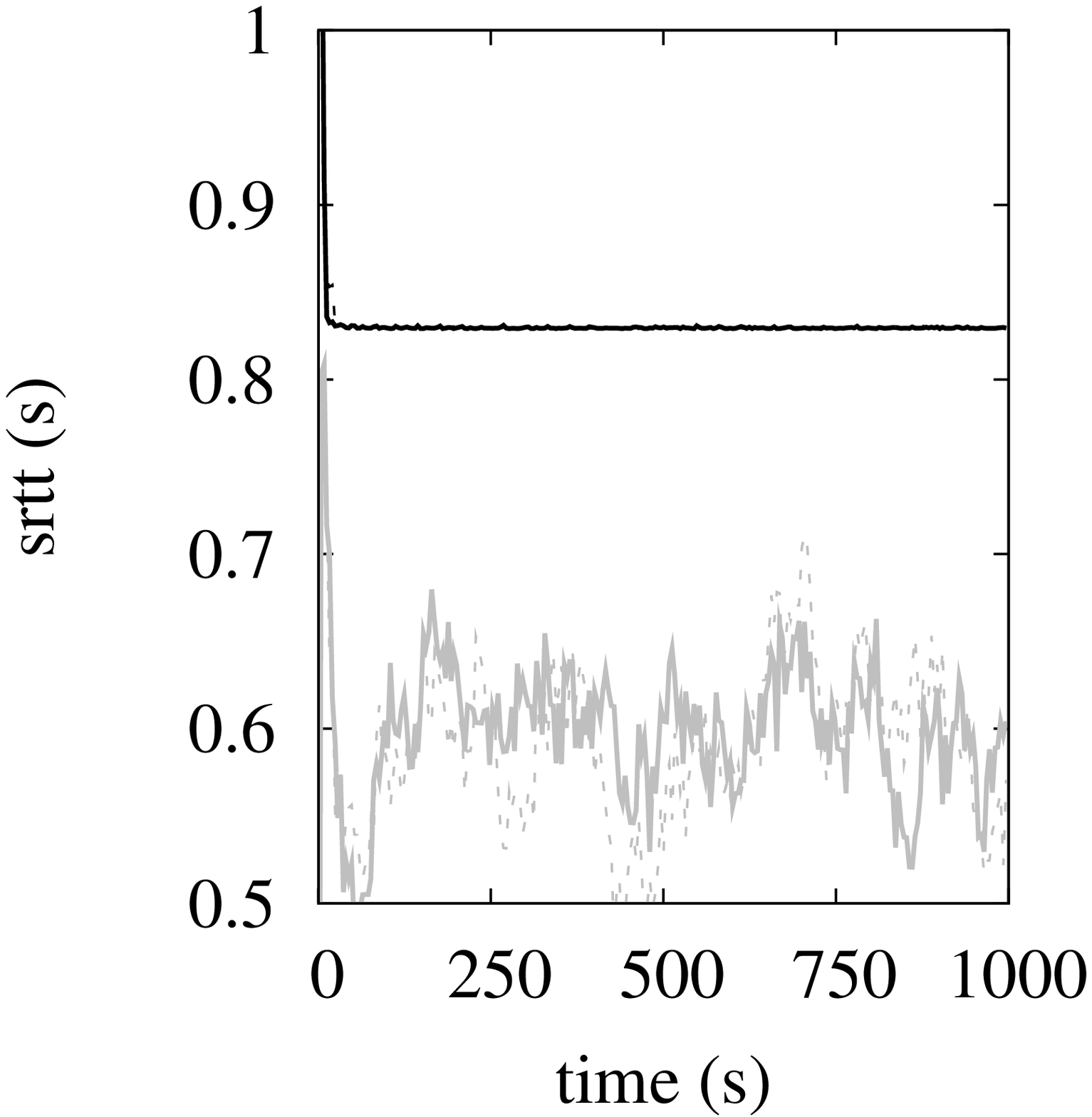}\label{fig:rtt_25}}
\subfloat[$p=0.1855$]{\includegraphics[width=0.19\textwidth]{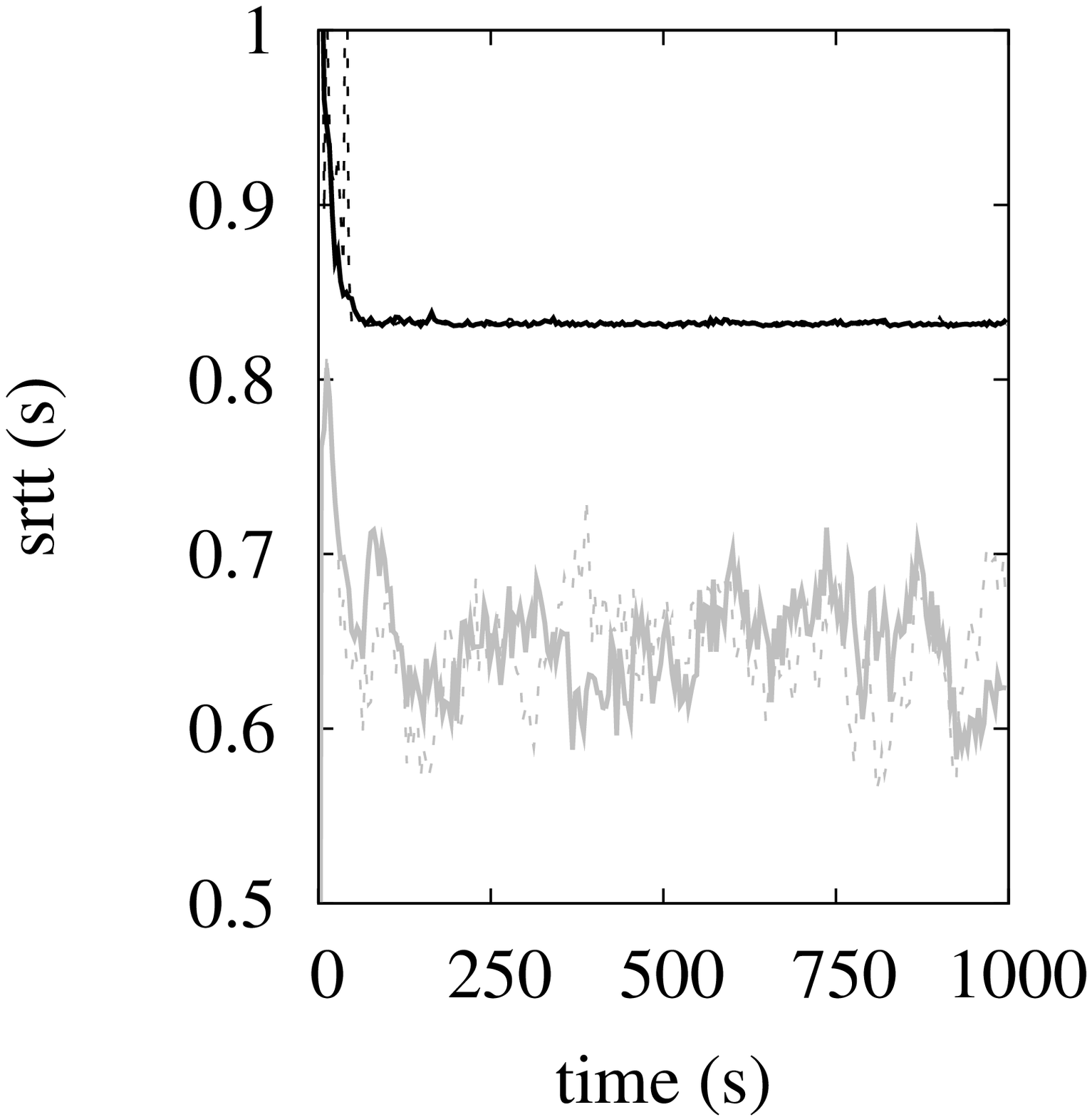}\label{fig:rtt_50}}
\subfloat{\includegraphics[width=0.065\textwidth]{legend}}
\end{center}\vspace*{-.2cm}\caption{The round trip time estimate (SRTT) of TCP/NC and TCP with varying link erasure probability $p$.}\label{fig:rtt}\vspace*{-.2cm}
\end{figure*}
\begin{figure*}[tbp]
\begin{center}
\subfloat[$R=1.10$]{\includegraphics[width=0.19\textwidth]{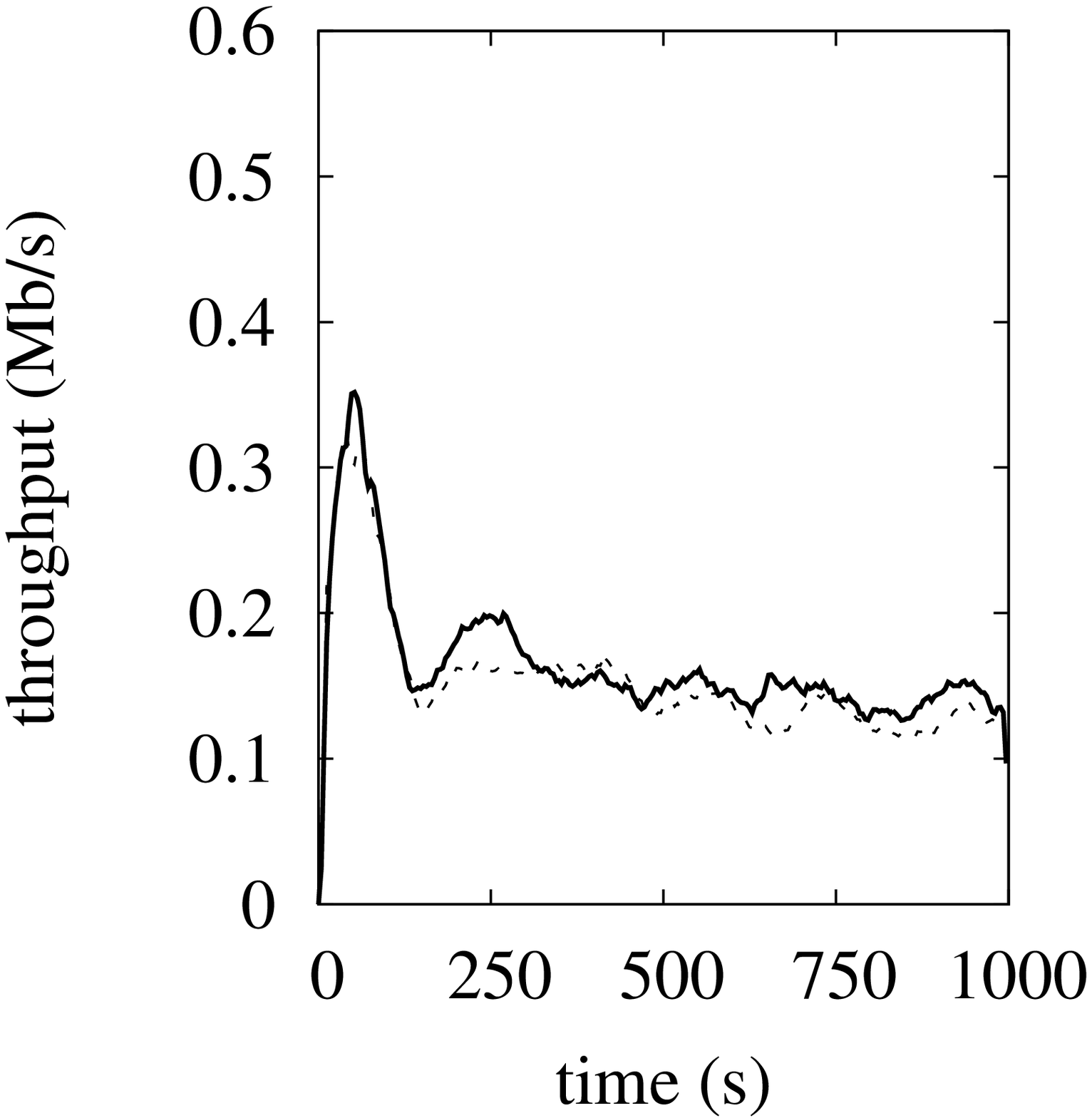}\label{fig:25bw_110}}
\subfloat[$R=1.11$]{\includegraphics[width=0.19\textwidth]{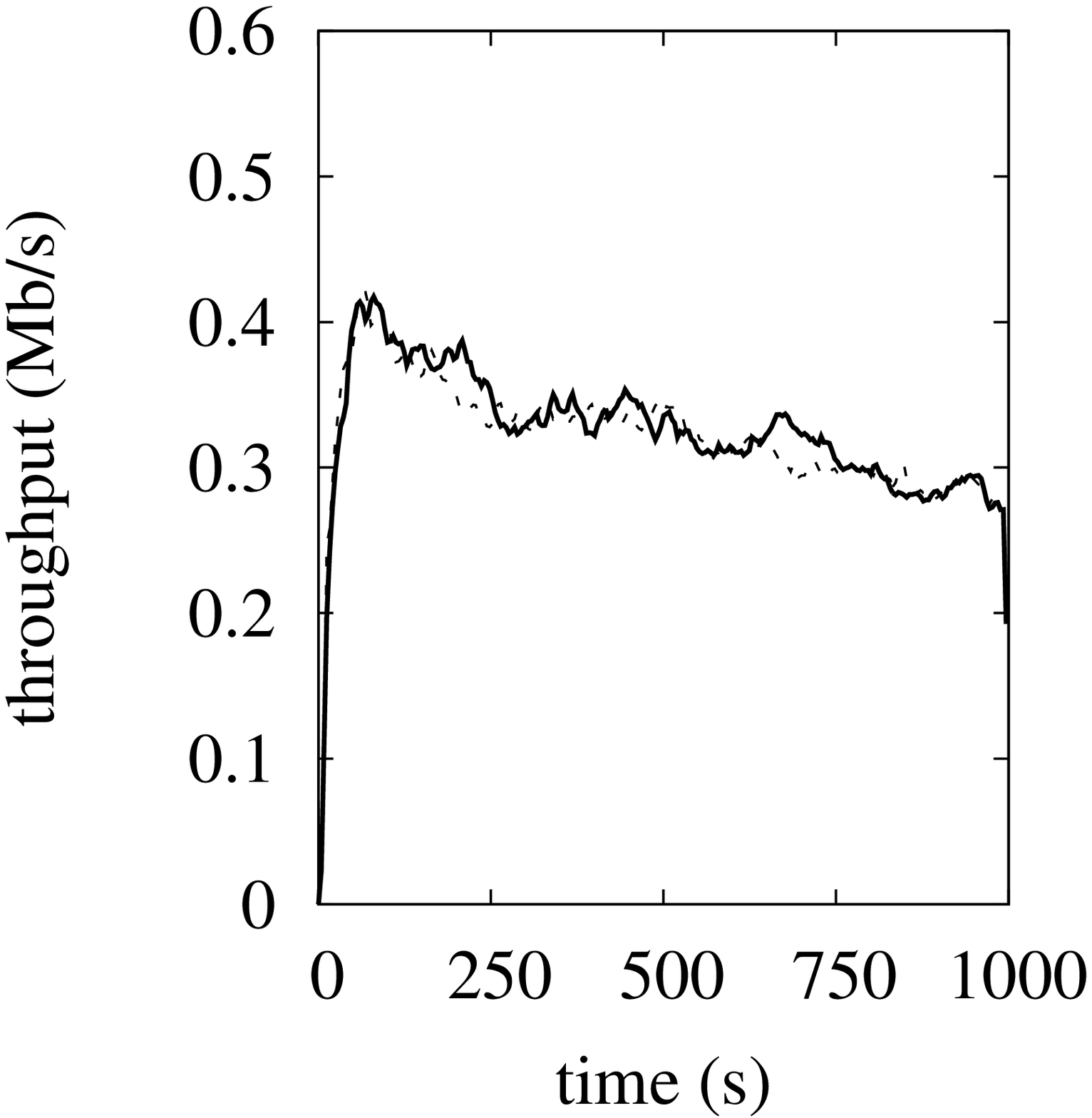}\label{fig:25bw_111}}
\subfloat[$R=1.12$]{\includegraphics[width=0.19\textwidth]{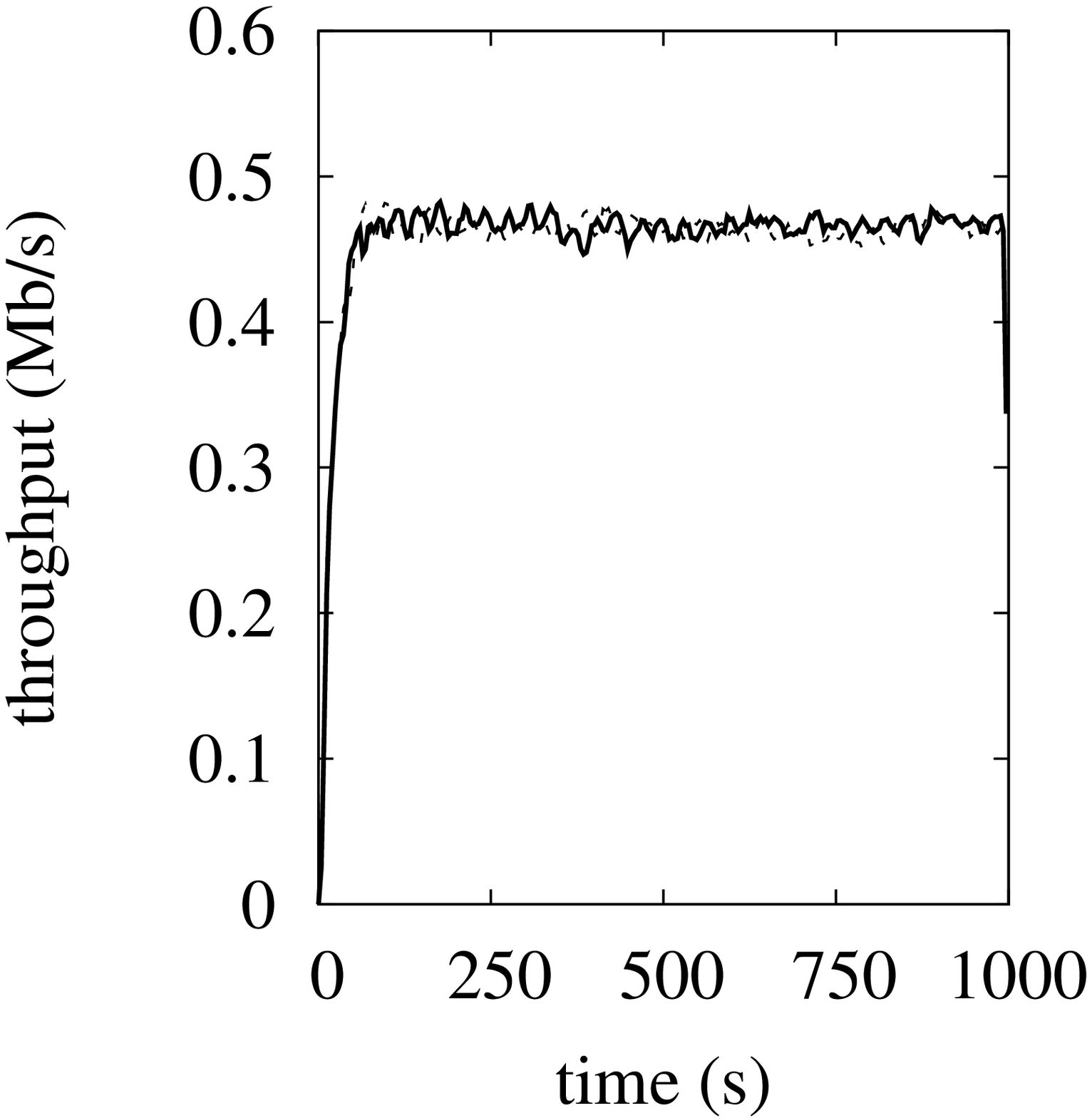}\label{fig:25bw_112}}
\subfloat[$R=1.13$]{\includegraphics[width=0.19\textwidth]{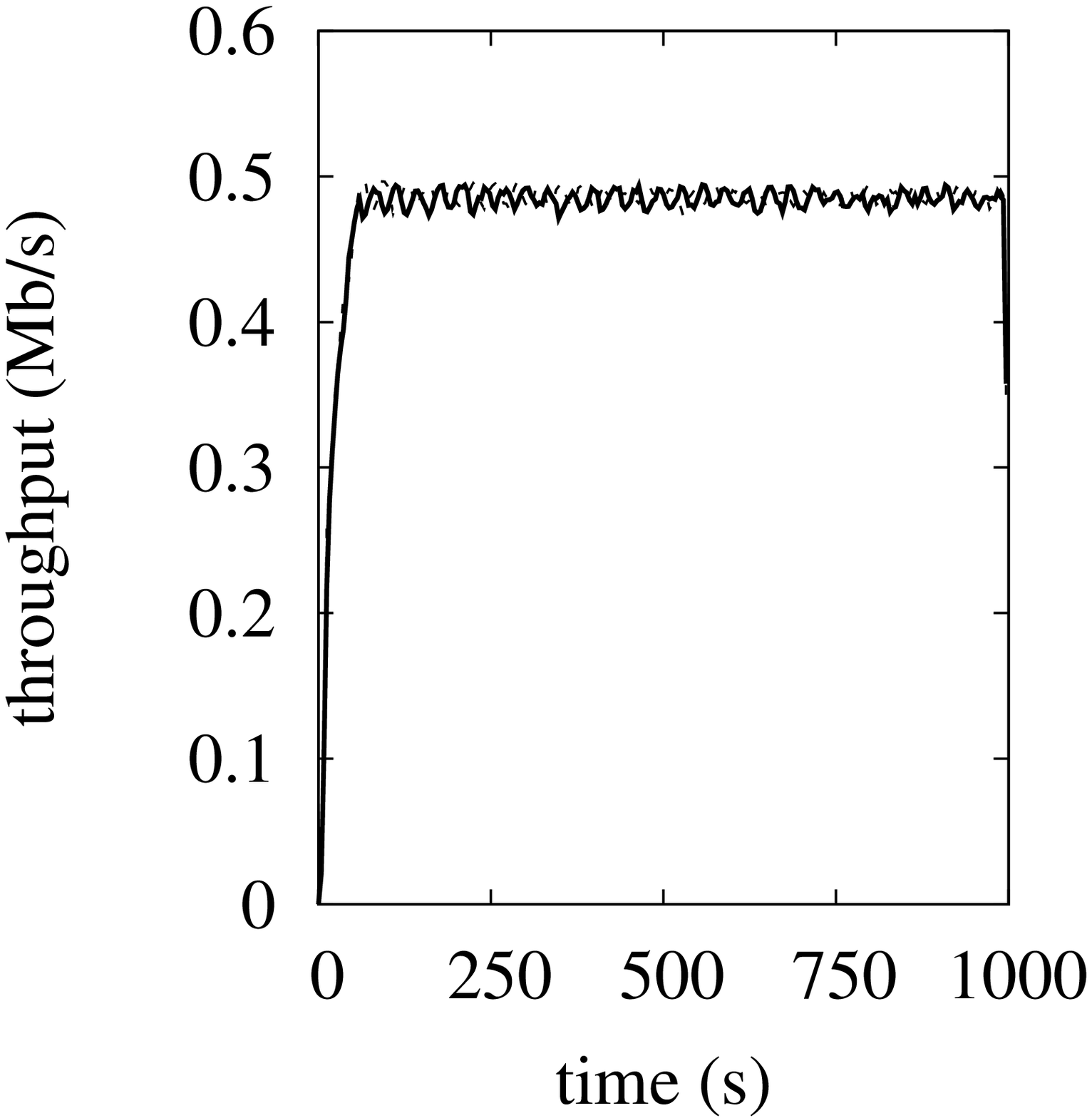}\label{fig:25bw_113}}
\subfloat[$R=1.15$]{\includegraphics[width=0.19\textwidth]{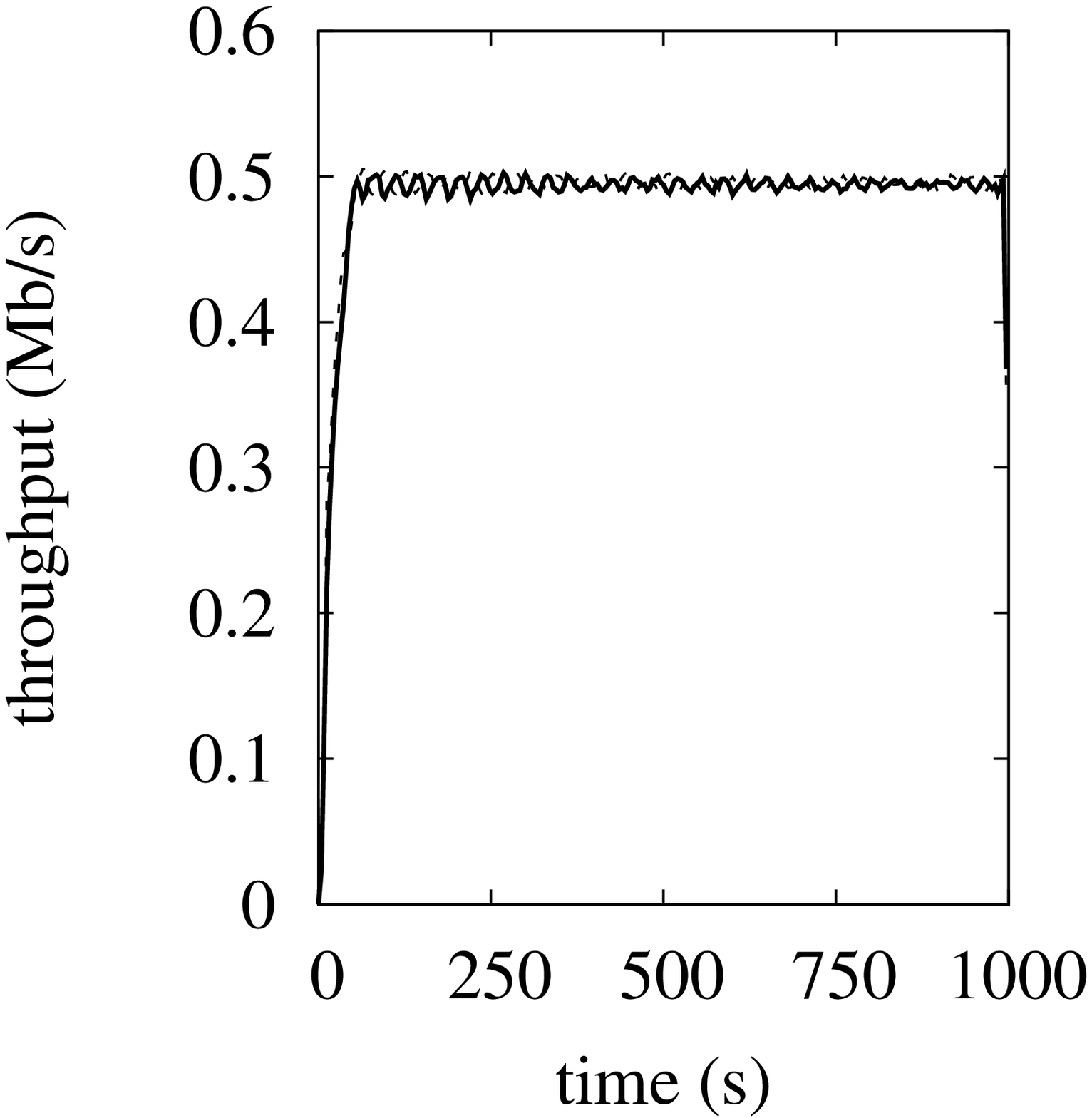}\label{fig:25bw_115}}
\subfloat{\includegraphics[width=0.065\textwidth]{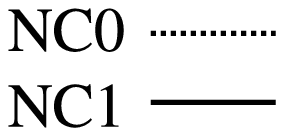}}
\end{center}\vspace*{-.2cm}\caption{Throughput of TCP/NC for $p=0.0963$ with varying redundancy factor $R$. Note that $\frac{1}{1-p} = 1.107$.}\label{fig:25bw}\vspace*{-.2cm}
\end{figure*}

The network topology for the simulation is shown in Figure \ref{fig:setup}. All links, in both forward and backward paths, are assumed to have a bandwidth of $C$ Mbps, a propagation delay of 100 ms, a buffer size of 200, and a erasure rate of $q$. Note that since there are in total four links in the path from node 0 to node 4, the probability of packet erasure is $p = 1-(1-q)^4$. Each packet transmitted is assumed to be 8000 bits (1000 bytes). We set $W_{\max}=50$ packets for all simulations. In addition, time-out period $T_o = \frac{3}{RTT} = 3.75$ rounds long (3 seconds). Therefore, our variables for the simulations are:
\begin{itemize}
\item $p = 1-(1-q)^4$: End-to-end erasure rate,
\item $R$: Redundancy factor,
\item $C$: Capacity of the links (in Mbps).
\end{itemize}
We study the effect these variables have on the following:
\begin{itemize}
\item $\thru{}$: Throughput of TCP or TCP/NC,
\item $E[W]$: Average window size of TCP or TCP/NC,
\item $SRTT$: Round-trip estimate.
\end{itemize}
For each data point, we average the performance over 100 independent runs of the simulation, each of which is 1000 seconds long.





\subsection{Probability of erasure $p$}\label{sec:prob}

We set $C = 2$ Mbps  and $R = 1.25$ regardless of the value of $p$. We vary $q$ to be 0, 0.005, 0.015, 0.025, and 0.05. The corresponding $p$ values are 0, 0.0199, 0.0587, 0.0963, and 0.1855. The 
results are shown in Figures \ref{fig:bw}, \ref{fig:cw}, and \ref{fig:rtt}.

Firstly, we show that when there are no random erasures ($p=0$), then TCP/NC and TCP behave similarly, as shown in Figures \ref{fig:bw_0}, \ref{fig:cw_0}, and \ref{fig:rtt_0}. Without any random losses and congestion, all of the flows (NC0, NC1, TCP0, TCP1) achieve the maximal throughput, $\frac{W_{\max}}{RTT} \cdot \frac{8}{10^6} =0.5$ Mbps.

The more interesting result is when $p>0$. 
As our analysis predicts, TCP/NC sustains its high throughput despite the random erasures in the network. We observe that TCP may close its window due to triple-duplicates ACKs or timeouts;
however, TCP/NC is more resilient to such erasure patterns. Therefore, TCP/NC is able to increment its window consistently, and \emph{maintain} the window size of 50 even under lossy conditions when standard TCP is unable to (resulting in the window fluctuation in Figure \ref{fig:cw}).

An interesting observation is that, TCP achieves a moderate average window size although the throughput (Mbps) is much lower (Figures \ref{fig:bw} and \ref{fig:cw}). This shows that na\"ively keeping the transmission window open is not sufficient to overcome the random losses within the network, and does not lead to improvements in TCP's performance. Even if the transmission window is kept open (e.g. during timeout period), the sender can not transmit additional packets into the network without receiving ACKs. Eventually, this leads to a TD or TO event.

As described in Sections \ref{sec:td-intuition} and \ref{sec:srtt}, TCP/NC masks errors by translating losses as longer RTT. For TCP/NC, if a specific packet is lost, the next subsequent packet received can ``replace'' the lost packet; thus, allowing the receiver to send an ACK. Therefore, the longer RTT estimate takes into account the delay associated with waiting for the next subsequent packet at the receiver. In Figure \ref{fig:rtt}, we verify that this is indeed true. TCP, depending on the ACKs received, modifies its RTT estimation; thus, due to random erasures, TCP's RTT estimate fluctuates significantly. On the other hand, TCP/NC is able to maintain a consistent estimate of the RTT; however, is slightly above the actual 800 ms.

\begin{figure}[tbp]
\begin{center}
\subfloat[$R=1.26$]{\includegraphics[width=0.19\textwidth]{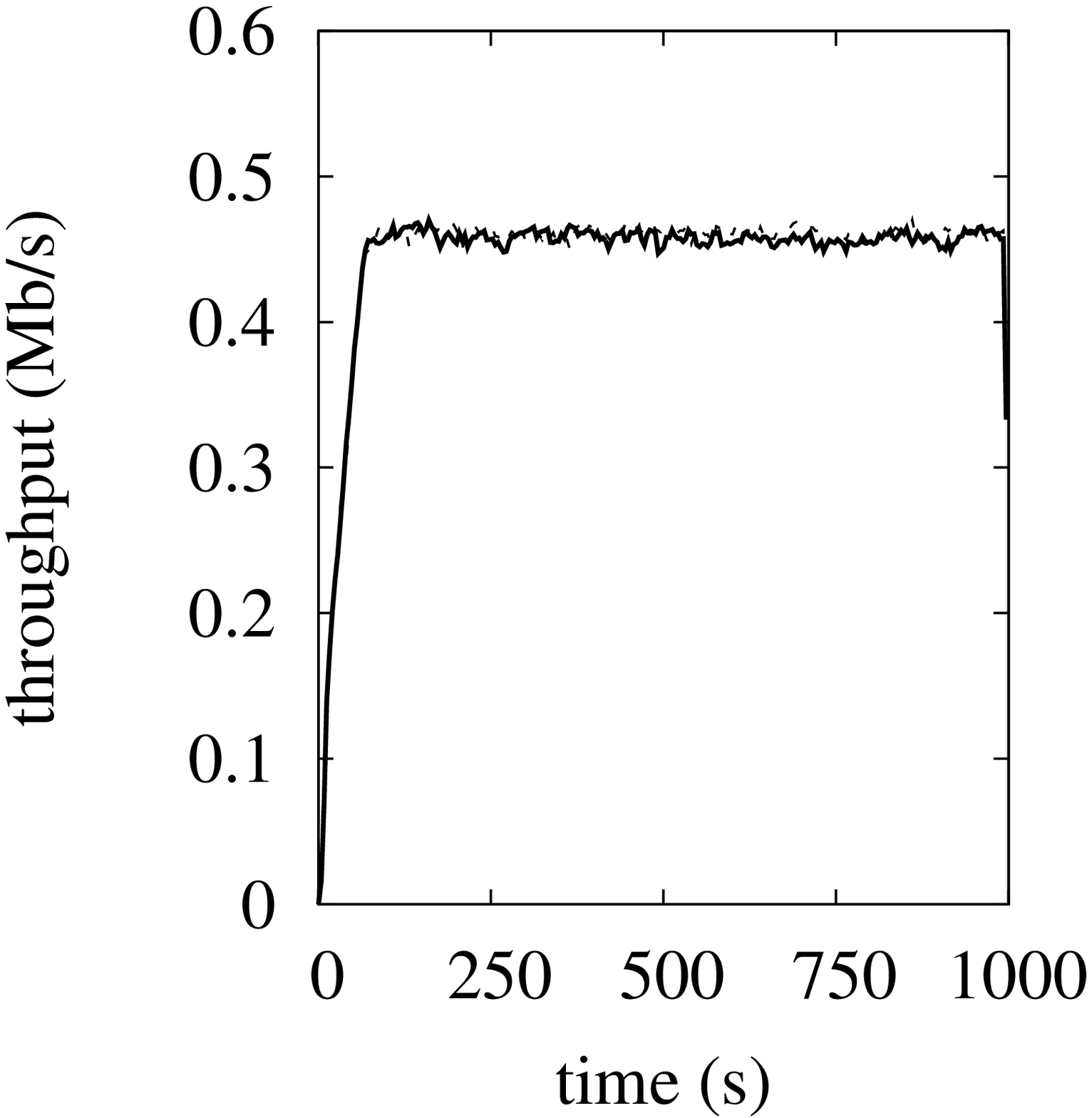}\label{fig:50bw_126}}
\subfloat[$R=1.28$]{\includegraphics[width=0.19\textwidth]{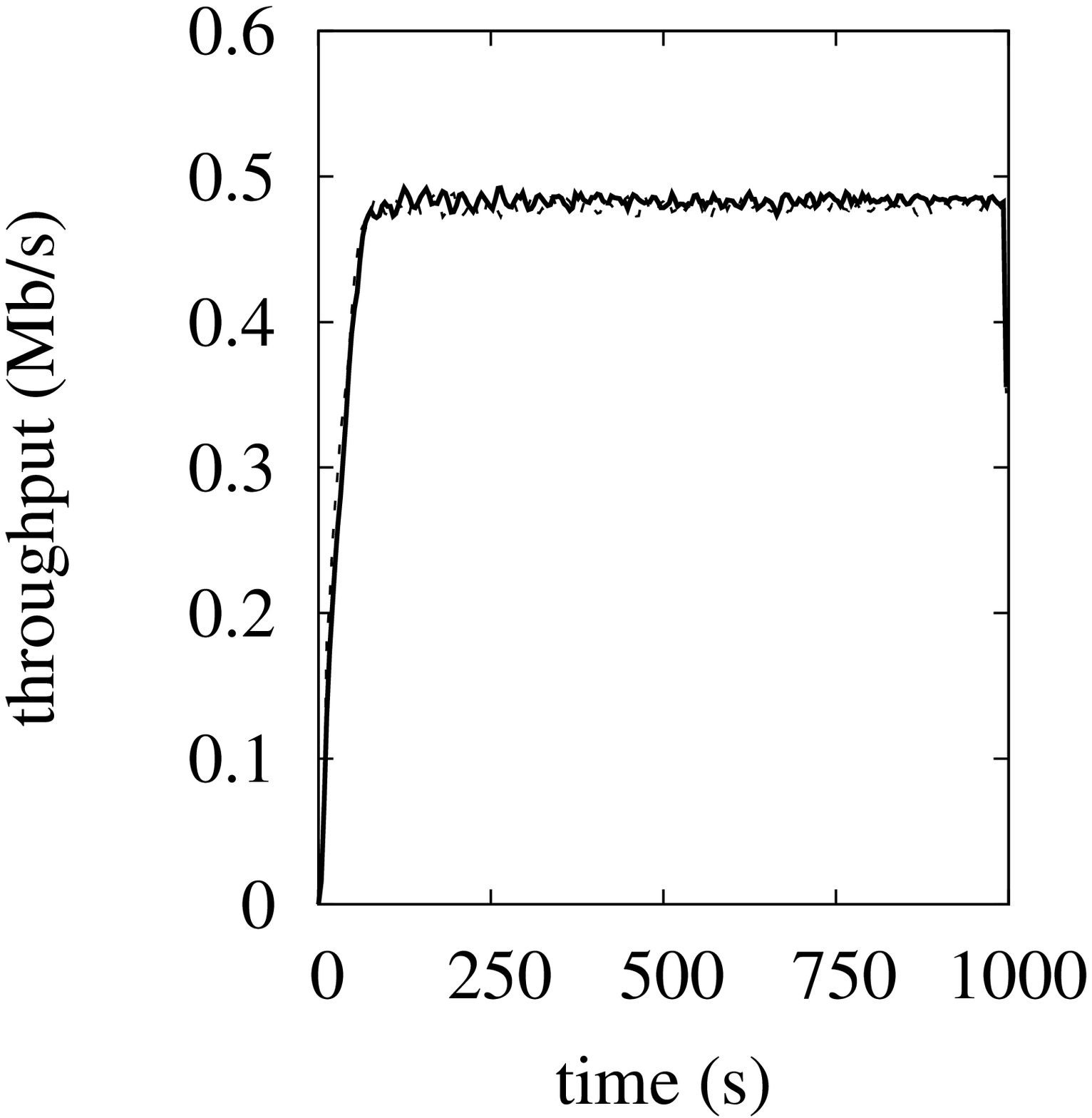}\label{fig:50bw_128}}
\subfloat{\includegraphics[width=0.065\textwidth]{legend_nc}}
\end{center}\vspace*{-.2cm}\caption{Throughput of TCP/NC for $p=0.1855$ with varying redundancy factor $R$. Note that $\frac{1}{1-p} = 1.228$.}\label{fig:50bw}\vspace*{-.2cm}
\end{figure}

\subsection{Redundancy factor $R$}\label{sec:redundancy_sim}

We set $C = 2$ Mbps. We vary the value of $p$ and $R$ to understand the relationship between $R$ and $p$. In Section \ref{sec:redundancy}, we noted that $R \geq \frac{1}{1-p}$ is necessary to mask random erasures from TCP. However, as $R \rightarrow \frac{1}{1-p}$, the probability that the erasures are completely masked decreases. This may suggest that we need $R \gg \frac{1}{1-p}$ for TCP/NC to maintain its high throughput. However, we shall show that $R$ need not be much larger than $\frac{1}{1-p}$ for TCP/NC to achieve its maximal throughput.

In Figure \ref{fig:25bw}, we present TCP/NC throughput behavior with $p = 0.0963$ and varying $R$. Note that $\frac{1}{1-p} = 1.107$ for $p = 0.0963$. There is a dramatic change in throughput behavior as we increase $R$ from 1.11 to 1.12. Note that $R = 1.12$ is only 1\% additional redundancy than the theoretical minimum, i.e. $\frac{1.12}{1/(1-p)} \approx 1.01$. Another interesting observation is that, even with $R = 1.10$ or $R=1.11$, TCP/NC achieves a significantly higher throughput than TCP (in Figure \ref{fig:bw_25}) for $p = 0.0963$.

Figure \ref{fig:bw_50} shows that, with $p = 0.1855$, TCP/NC throughput is not as steady, and does not achieve the maximal throughput of 0.5 Mbps.
This is because $\frac{1}{1-p} = 1.23$ is very close to $R=1.25$. As a result, $R = 1.25$ is not sufficient to mask erasures with high probability. In Figure \ref{fig:50bw}, we show that TCP/NC achieves an average throughput of 0.5 Mbps once $R\geq 1.28$. Note that $R = 1.28$ is only 4\% additional redundancy than the theoretical minimum, i.e. $\frac{1.28}{1/(1-p)} \approx 1.04$.

Similar behavior can be observed for $p = 0.0199$ and $0.0587$, and setting $R$ to be slightly above $\frac{1}{1-p}$ is sufficient.
A good heuristic to use in setting $R$ is the following. Given a probability of erasure $p$ and window size $W$, the probability that losses in any given round is completely masked is upper bounded by $\sum_{x=0}^{W(R-1)} \binom{RW}{x} p^x (1-p)^{RW-x}$, i.e. there are no more than $W(R-1)$ losses in a round. Ensuring that this probability is at least 0.8 has proven to be a good heuristic to use in finding the appropriate value of $R$.

\begin{figure*}[tbp]
\begin{minipage}[t]{5cm}
\begin{center}
\subfloat{\includegraphics[width=0.68\textwidth]{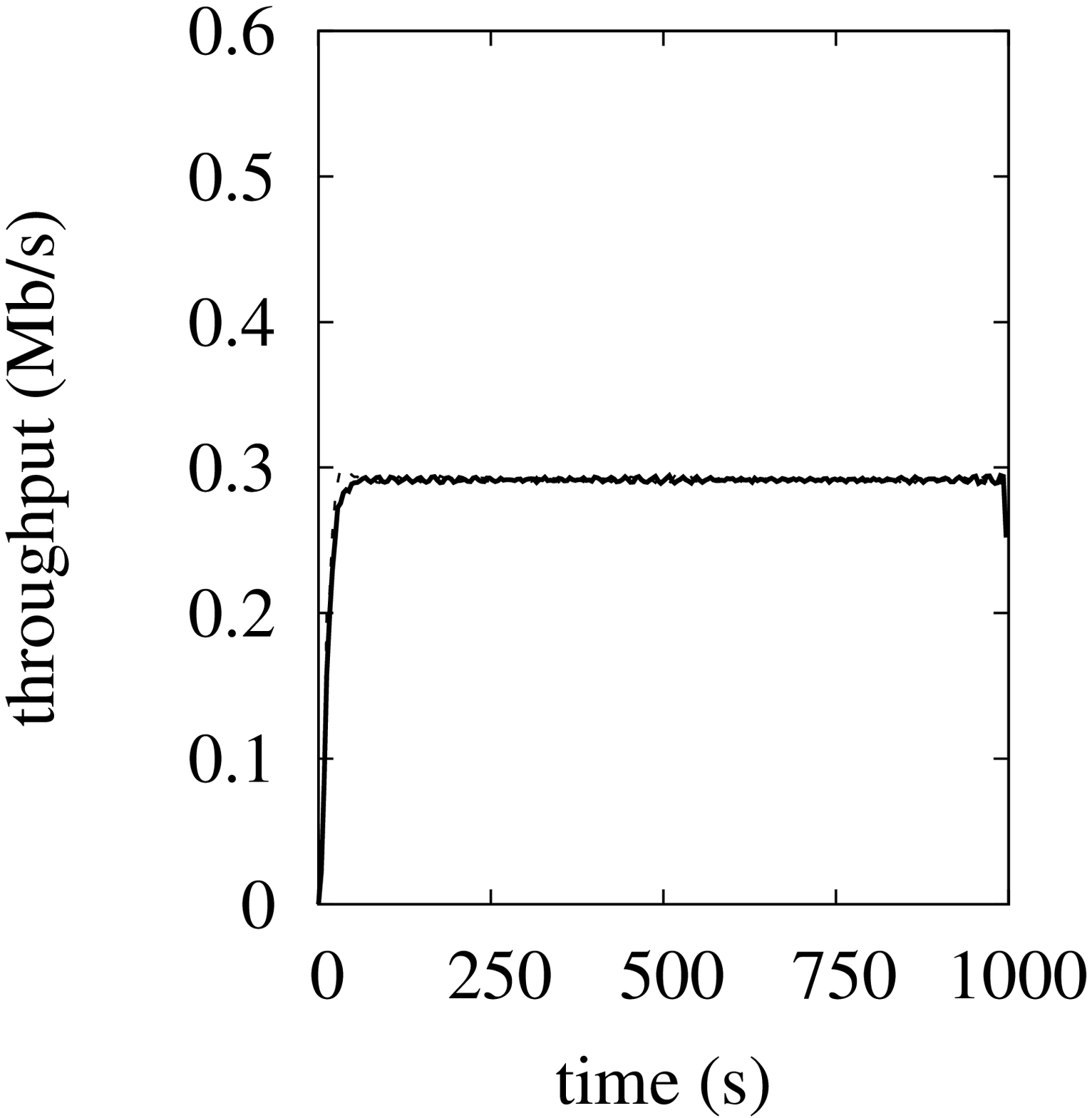}}
\subfloat{\includegraphics[width=0.24\textwidth]{legend_nc}}
\end{center}\vspace*{-.2cm}\caption{TCP/NC for $p=0.0963$ and $C=0.7$ Mbps.}\label{fig:congestion_2}
\end{minipage}
\hfill
\begin{minipage}[t]{12cm}
\begin{center}
\subfloat{\includegraphics[width=0.28\textwidth]{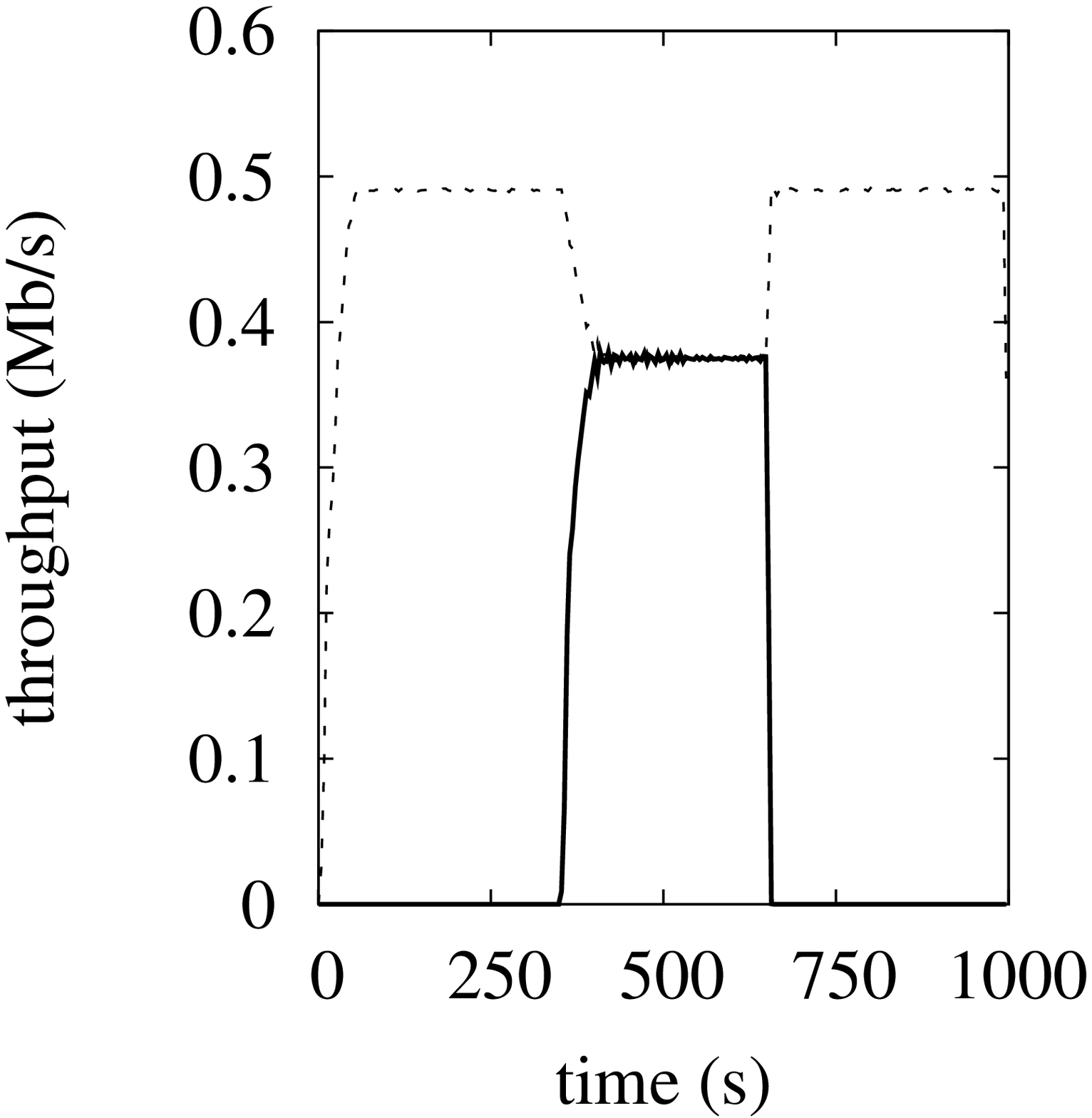}\label{fig:bw_congestion}}
\subfloat{\includegraphics[width=0.28\textwidth]{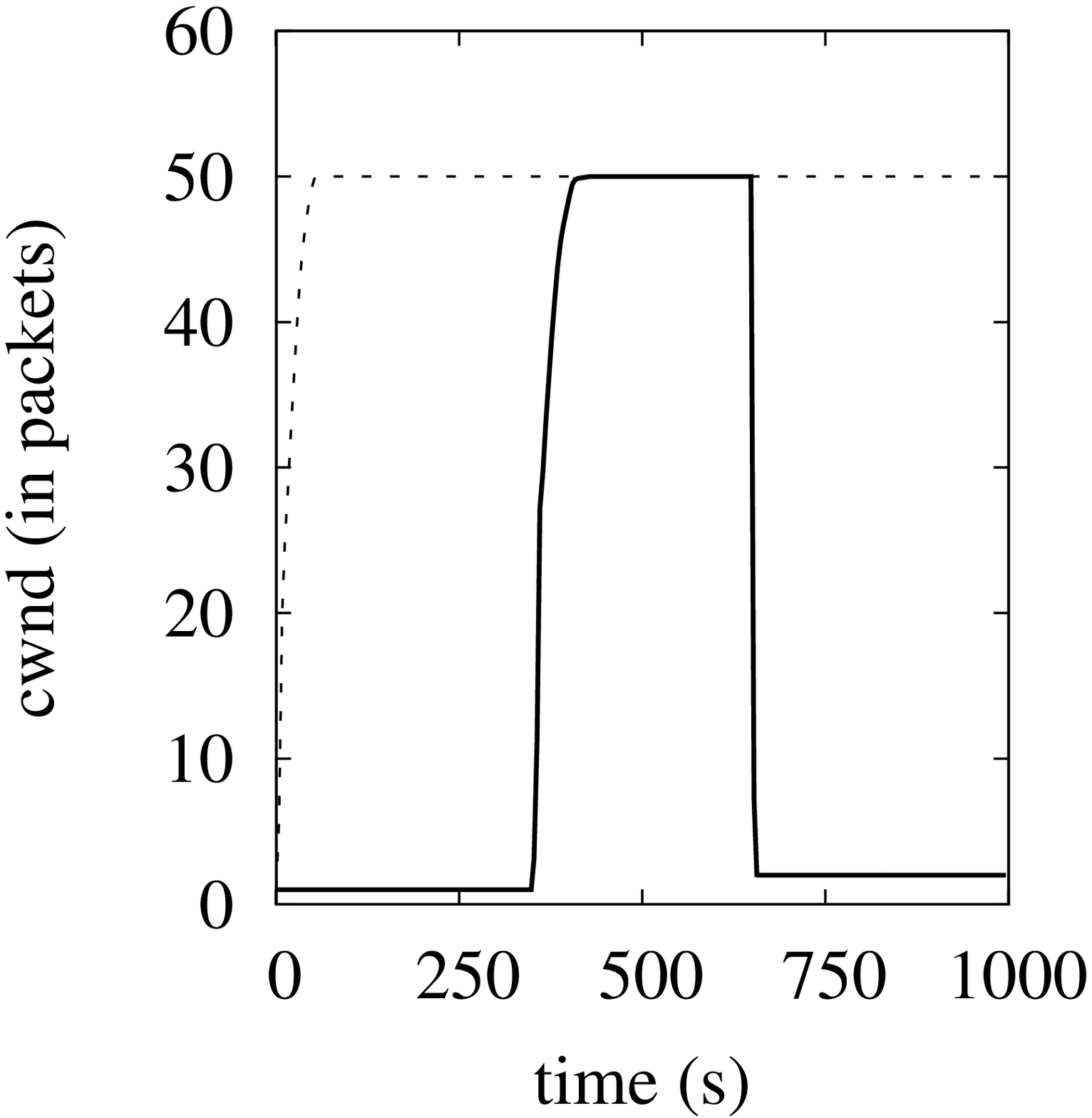}\label{fig:cw_congestion}}
\subfloat{\includegraphics[width=0.28\textwidth]{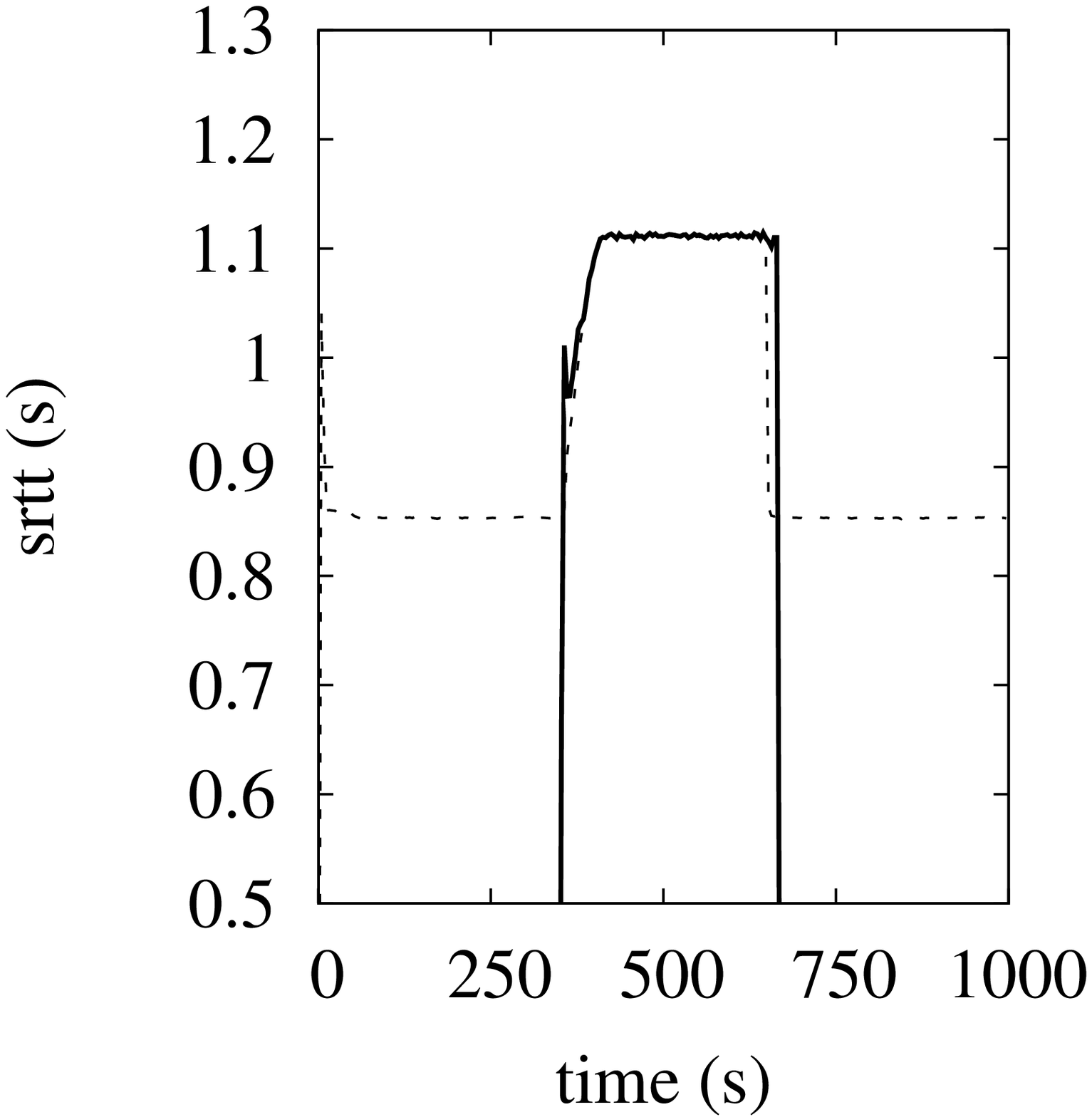}\label{fig:rtt_congestion}}
\subfloat{\includegraphics[width=0.1\textwidth]{legend_nc}}
\end{center}\vspace*{-.2cm}\caption{TCP/NC-TCP/NC for $p=0.0963$ with congestion ($C = 0.9$ Mbps, $R=1.2$, $W_{\max}=50$).}\label{fig:congestion}
\end{minipage}
\end{figure*}

\begin{table*}[tbp]
\caption{The average simulated or predicted long-term throughput of TCP and TCP/NC in megabits per second (Mbps). `NC0', 'NC1', `TCP0', `TCP1' are average throughput achieved in the NS-2 simulations (with the corresponding `$R$'). `TCP/NC analysis' is calculated using Equation (\ref{eq:nc-final-longterm}) with $\lfloor n\cdot SRTT \rfloor = 1000$. `TCP analysis' is computed using Equation (\ref{eq:tcp}). }\label{tb:thru}
\centering
\begin{tabular}{|c|c|c|c|c|c|c|c|c|}
\hline
$p$ & TCP/NC $SRTT$ & $R$ & NC0 & NC1 & TCP/NC analysis & TCP0 & TCP1 & TCP analysis \\
\hline
 0      & 0.8256 & 1 & 0.5080 & 0.5057 & 0.4819 & 0.5080 & 0.5057 & 0.5000 \\
 0.0199 & 0.8260 & 1.03 & 0.4952 & 0.4932 & 0.4817 & 0.1716 & 0.1711 & 0.0667\\
 0.0587 & 0.8264 & 1.09 & 0.4926 & 0.4909 & 0.4814 & 0.0297 & 0.0298 & 0.0325\\
 0.0963 & 0.8281 & 1.13 & 0.4758 & 0.4738 & 0.4804 & 0.0149 & 0.0149 & 0.0220\\
 0.1855 & 0.8347 & 1.29 & 0.4716 & 0.4782 & 0.4766 & 0.0070 & 0.0070 & 0.0098\\
\hline
\end{tabular}
\end{table*}

\subsection{Congestion Control}\label{sec:congestioncontrol}

We showed that TCP/NC achieves a good performance in lossy environment. This may raise concerns about masking correlated losses from TCP; thus, disabling TCP's congestion control mechanism. We show that the network coding layer masks random losses only, and allows TCP's congestion control to take affect when necessary.

Given a capacity $C$ and erasure rate $p$, the available bandwidth is $C(1-p)$ Mbps. Given two flows, a fair allocation of bandwidth should be $\frac{C(1-p)}{2}$ Mbps per flow. Note that this is the \emph{available} bandwidth, not the \emph{achieved} bandwidth.
As we have seen, if $p >0$, TCP may not be able to use fully the available bandwidth. On the other hand, TCP/NC is able to use the available bandwidth efficiently.
With TCP/NC flows, there is another parameter we need to consider: the redundancy factor $R$. Since TCP/NC sends $R$ coded packets for each data packet, the achievable bandwidth is $\min\{C(1-p),\frac{C}{R}\}$ Mbps; if shared among two flows fairly, we expect $\frac{1}{2} \min\{C(1-p),\frac{C}{R}\}$ Mbps per coded flow. Note that, if $R$ is chosen appropriately (i.e. slightly above $\frac{1}{1-p}$), then TCP/NC can achieve rate close to $C(1-p)$, which is optimal.

We show that multiple TCP/NC flows share the bandwidth fairly. We consider two flows (NC0, NC1) with $W_{\max}=50$, $R =1.2$, and $p=0.0963$. If there is no congestion, each flow would achieve approximately 0.5 Mbps. However, we set $C=0.7$ Mbps. The two flows should achieve $\frac{1}{2} \min\{0.7(1-0.0963),\frac{0.7}{1.2}\} = 0.2917$ Mbps. We observe in Figure \ref{fig:congestion_2} that NC0 and NC1 achieve 0.2878 Mbps and 0.2868 Mbps, respectively. Note that $\frac{C(1-p)}{2} = 0.3162$; thus, NC0 and NC1 is near optimal even though $R = 1.2 > \frac{1}{1-p}=1.106$.

For our next simulations, we set $C = 0.9$ Mbps, $W_{\max}=50$, $p = 0.0963$, and $R = 1.2$. Furthermore, we assume that NC0 starts at 0s, and runs for 1000s, while NC1 starts at time 350s and ends at time 650s. 
Before NC1 enters, NC0 should be able to achieve a throughput of 0.5 Mbps; however, when NC1 starts its connection, there is congestion, and both NC0 and NC1 have to react to this. Figure \ref{fig:congestion} shows that indeed this is true. We observe that when NC1 starts its connection, both NC0 and NC1 shares the bandwidth equally (0.3700 and 0.3669 Mbps, respectively). The achievable bandwidth predicted by $\min\{C(1-p),\frac{C}{R}\}$ is 0.75 Mbps (or 0.375 Mbps per flow). Note that both NC0 and NC1 maintains its maximum window size of 50. Instead, NC0 and NC1 experience a longer RTT, which naturally translates to a lower throughput given the same $W_{\max}$.

\subsection{Comparison to the analytical model}\label{sec:compare}

Finally, we examine the accuracy of our analytical model in predicting the behavior of TCP and TCP/NC. First, note that our analytical model of window evolution (shown in Equation (\ref{eq:tcpnc-w2}) and Figure \ref{fig:period}) demonstrates the same trend as that of the window evolution of TCP/NC NS-2 simulations (shown in Figure \ref{fig:cw}). Second, we compare the actual NS-2 simulation performance to the analytical model. This is shown in Table \ref{tb:thru}. We observe that Equations (\ref{eq:nc-throughput-round}) and (\ref{eq:tcpnc-w2}) predict well the trend of TCP/NC's throughput and window evolution, and provides a good estimate of TCP/NC's performance. Furthermore, our analysis predicts the average TCP behavior well. In Table \ref{tb:thru}, we see that Equation (\ref{eq:tcp}) is consistent with the NS-2 simulation results even for large values of $p$. Therefore, both simulations as well as analysis support that TCP/NC is resilient to erasures; thus, better suited for reliable transmission over unreliable networks, such as wireless networks.

\section{Model for Network Cost}\label{sec:bs_model}

Mobile data traffic has been growing at an alarming rate with some estimating that it will increase more than 25-folds in the next five years \cite{cisco}. In order to meet such growth, there has been an increasing effort to install and upgrade the current networks. As shown in Figure \ref{fig:towers}, mobile service providers often install more infrastructure (e.g. more base stations) in areas which already have full coverage. The new infrastructure is to provide more bandwidth, which would lead to higher quality of experience to users. However, this increase in bandwidth comes at a significant energy cost as each base station has been shown to use 2-3 kilowatts (kW) \cite{bell_power}. The sustainability and the feasibility of such rapid development have been brought to question as several trends indicate that the technology efficiency improvements may not be able to keep pace with the traffic growth \cite{bell_power}.

In the subsequent sections, we use the results from Sections \ref{sec:td}, \ref{sec:tcpnc}, and \ref{sec:simulations} to show that TCP/NC allows a better use of the base stations installed and can improve the goodput without any additional base stations. Improving the goodput with the same or a fewer number of base stations implies reduction in energy cost, operational expenses, capital expenses, and maintenance cost for the network provider. The results in this paper can also be understood as being able to serve more users or traffic growth with the same number of base stations. This may lead to significant cost savings, and may be of interest for further investigation.

\subsection{Model}\label{sec:bs_setup}

\begin{figure}[tbp]
\begin{center}\vspace*{.4cm}
\includegraphics[width=.35\textwidth]{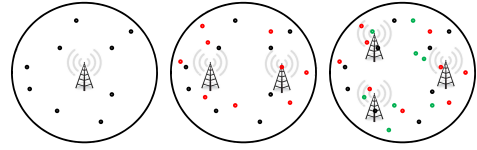}
\end{center}\vspace*{-.2cm}\caption{As number of users in a given area grows, a service provider may add additional base stations not for coverage but for bandwidth. As red users join the network, a second base station may be necessary; as green users join the network, a third base station may become necessary in order to maintain a certain level of quality of service.
}\label{fig:towers}\vspace*{-.2cm}
\end{figure}

Consider a network with $N$ users. We assume that these $N$ users are in an area such that a single base station can cover them as shown in Figure \ref{fig:towers}. If the users are far apart enough that a single base station cannot cover the area, then more base stations are necessary; however, we do not consider the problem of coverage.

The network provider's goal is to provide a \emph{fair} service to any user that wishes to start a transaction. Here, by fair, we mean that \emph{every user is expected to be allocated the same average bandwidth}, denoted as $\band$ Mbps. The network provider wishes to have enough network resources, measured in number of base stations, so that any user that wishes to start a transaction is able to join the network immediately and is given an average bandwidth of $\band$ Mbps. We denote $\thru{}$ to be the throughput seen by the user. Note that $\thru{} \leq \band$.

We denote $N_{bs}$ to be the number of base stations needed to meet the network provider's goal. We assume that every base station can support at most $\band_{\max}$ Mbps (in bandwidth) and at most $N_{\max}$ active users simultaneously. In this paper, we assume that $\band_{\max} = 300$ Mbps and $N_{\max} = 200$. As previously, we denote $p$ to be the probability of packet loss in the network, and $RTT$ to be the round-trip time. 

A user is \emph{active} if the user is currently downloading a file; \emph{idle} otherwise. A user decides to initiate a transaction with probability $q$ at each time slot. Once a user decides to initiate a transaction, a file size of $f$ bits is chosen randomly according to a probability distribution $Q_f$. We denote $\mu_f$ to be the expected file size, and the expected duration of the transaction to be $\Delta = \mu_f/\thru{}$ seconds. If the user is already active, then the new transaction is added to the user's queue. If the user has initiated $k$ transactions, the model of adding the jobs into the user's queue is equivalent to splitting the throughput $\thru{}$ to $k$ transactions (each transaction achieves a throughput of $\thru{}/k$ Mbps).

\section{Analysis of the Number of Base Stations}\label{sec:bs_analysis}
We analyze $N_{bs}$ needed to support $N$ users given bandwidth $\band$ and throughput $\thru{}$. We first analyze $P(\Delta, q)$, the probability that a user is active at any given point in time. Given $P(\Delta, q)$, we compute the expected number of active users at any given point in time and $N_{bs}$ needed to support these active users.


To derive $P(\Delta, q)$, we use the Little's Law. For a stable system, the Little's Law states that the average number of jobs (or transactions in our case) in the user's queue is equal to the product of the arrival rate $q$ and the average transaction time $\Delta$. When $\Delta p \geq 1$, we expect the user's queue to have on average at least one transaction in the long run. This implies that the user is expected to be active at all times. When $\Delta p < 1$, we can interpret the result from Little's Law to represent the probability that a user is active. For example, if $\Delta p = 0.3$, the user's queue is expected to have 0.3 transactions at any given point in time. This can be understood as the user being active for 0.3 fraction of the time. Note that when the system is unstable, the long term average number of uncompleted jobs in the user's queue may grow unboundedly. 
In an unstable system, we assume that in the long term, a user is active with probability equal to one.

Therefore, we can state the following result for $P(\Delta, q)$.
\begin{equation}\label{eq:probability}
P(\Delta, q) = \min\lbrace1, \Delta q\rbrace = \min\left\lbrace1, \frac{\mu_f}{\thru{}}\cdot q\right\rbrace.
\end{equation}

Given $P(\Delta, q)$, the expected number of active users is $N P(\Delta, q)$. We can now characterize the expected number of base stations needed as
\begin{equation}\label{eq:basestations}
N_{bs} = NP(\Delta, q) \cdot \max{ \left\lbrace \frac{\band}{\band_{\max}}, \frac{1}{N_{\max}} \right\rbrace }.
\end{equation}
In Equation \eqref{eq:basestations}, $\max{ \lbrace \frac{\band}{\band_{\max}}, \frac{1}{N_{\max}} \rbrace }$ represents the amount of base stations' resources (the maximum load $\band_{\max}$ or the amount of activity $N_{\max}$) each active user consumes. The value of $N_{bs}$ from Equation \eqref{eq:basestations} may be fractional, indicating that actually $\lceil n_{bs}\rceil$ base stations are needed.

Note the effect of $\band$ and $\thru{}$. As shown in Equation \eqref{eq:basestations}, increasing $\band$ incurs higher cost while increasing $\thru{}$ reduces the cost. Therefore, when a network provider dedicates resources to increase $\band$, the goal of the network provider is to increase $\thru{}$ proportional to $\band$.

\section{Best Case Scenario}\label{sec:bestcase}

In an ideal scenario, the user should see a throughput $\thru{} = \band$. In this section, we analyze this best case scnario with  $\thru{} = \band$. This assumption can be considered as ignoring the effect of losses in the network; thus, TCP or TCP/NC can achieve the full bandwidth available. Once we understand the optimal scenario, we then consider the behavior of TCP and TCP/NC in Section \ref{sec:bs_tcpnc}.

\subsection{Analytical Results}\label{sec:analysisresults}

\begin{figure}[tbp]
\begin{center}
\subfloat[$\mu_f$ = 3.2 MB]{\includegraphics[width=.23\textwidth]{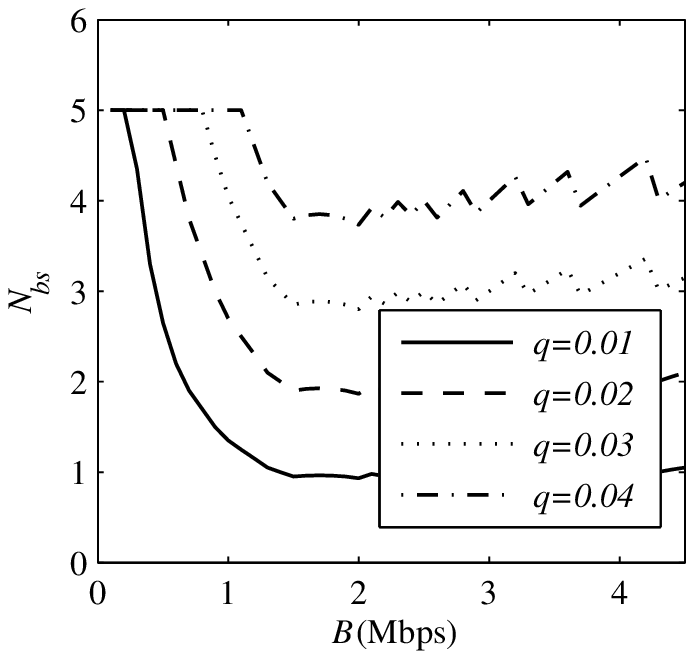}\label{fig:analysis_2}}
\subfloat[$\mu_f$ = 5.08 MB]{\includegraphics[width=.23\textwidth]{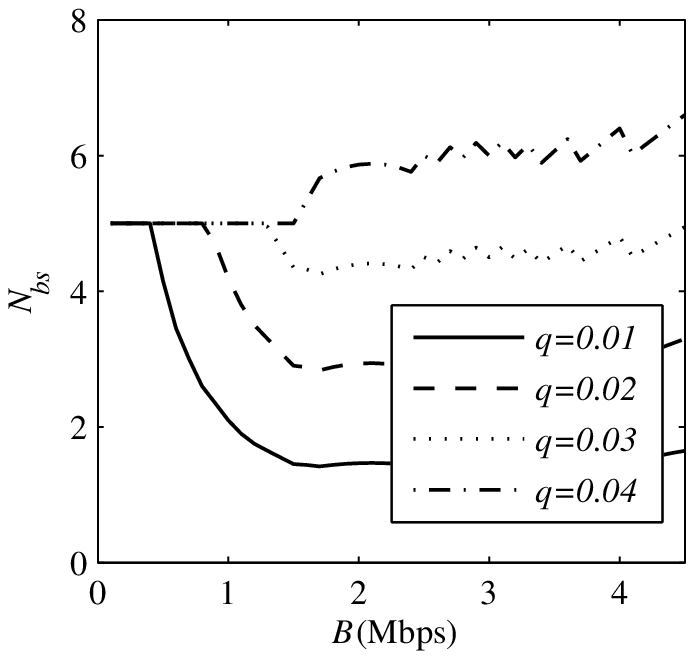}\label{fig:analysis_1}}
\end{center}\vspace*{-.2cm}
\caption{The values of $N_{bs}$ from Equation \eqref{eq:basestations} with $N=1000$ and varying $q$ and $\band$.}\vspace*{-.2cm}
\end{figure}

In Figures \ref{fig:analysis_2} and \ref{fig:analysis_1}, we plot Equation \eqref{eq:basestations} with $\mu_f = 3.2$ MB and $\mu_f = 5.08$ MB for varying values of $q$. As $\band$ increases, it does not necessarily lead to increase in $N_{bs}$. Higher $\band$ results in users finishing their transactions faster, which in turn allows the resources dedicated to these users to be released to serve other requests or transactions. As a result, counter-intuitively, we may be able to maintain a higher $\band$ with \emph{the same or a fewer} number of base stations than we would have needed for a lower $\band$.
For example, in Figure \ref{fig:analysis_2}, when $\band < 1$ Mbps, the rate of new requests exceeds the rate at which the requests are handled; resulting in an unstable system. As a result, most users are active all the time, and the system needs $\frac{n}{N_{\max}} = \frac{1000}{200} = 5$ base stations.

There are many cases where $N_{bs}$ is relatively constant regardless of $\band$. For instance, consider $q=0.03$ in Figure \ref{fig:analysis_1}. The value of $N_{bs}$ is approximately 4-5 throughout. However, there is a significant difference in the way the resources are used. When $\band$ is low, all users have slow connections; therefore, the base stations are fully occupied not in bandwidth but in the number of active users. On the other hand, when $\band$ is high, the base stations are being used at full-capacity in terms of bandwidth. As a result, although the system requires the same number of base stations, users experience better quality of service and users' requests are completed quickly.

When $q$ and $\band$ are high enough, it is necessary to increase $N_{bs}$. As demand exceeds the network capacity, it becomes necessary to add more infrastructure to meet the growth in demand. For example, consider $q = 0.04$ in Figure \ref{fig:analysis_1}. In this case, as $\band$ increases $N_{bs}$ increases.


\subsection{Simulation Results}
\begin{figure}[tbp]
\begin{center}
\subfloat[$\mu_f$ = 3.2 MB]{\includegraphics[width=.23\textwidth]{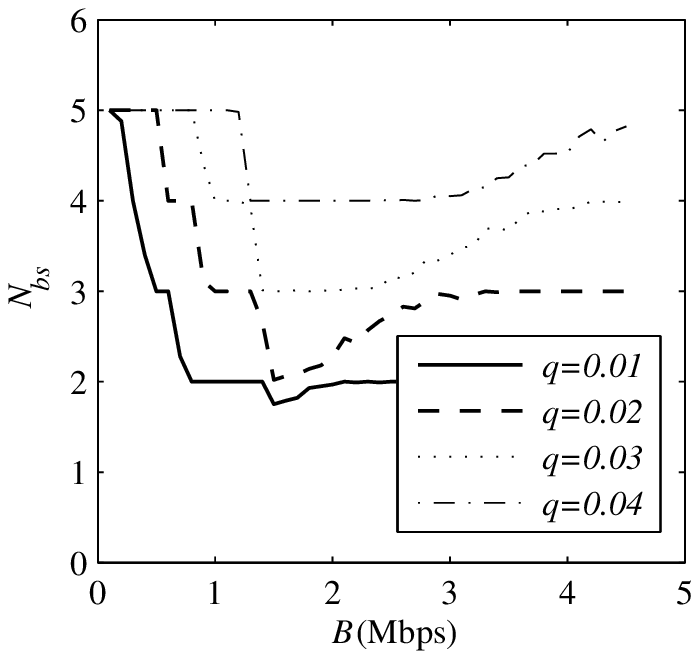}\label{fig:simulations_2}}
\subfloat[$\mu_f$ = 5.08 MB]{\includegraphics[width=.23\textwidth]{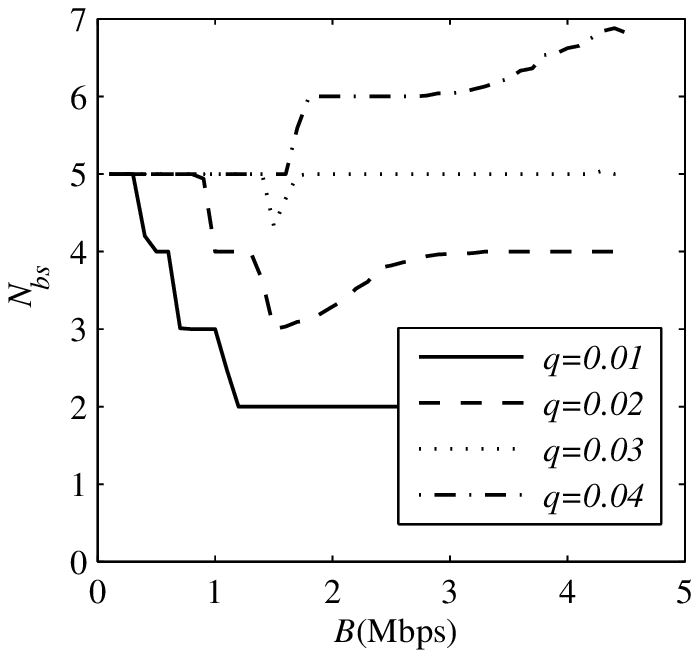}\label{fig:simulations_1}}
\end{center}\vspace*{-.2cm}
\caption{Average value of $N_{bs}$ over 100 iterations with $N=1000$ and varying $q$ and $\band$.}\vspace*{-.2cm}
\end{figure}

We present MATLAB simulation results to verify our analysis results in Section \ref{sec:analysisresults}. We assume that at every 0.1 second, a user may start a new transaction with probability $\frac{q}{10}$. This was done to give a finer granularity in the simulations; the results from this setup is equivalent to having users start a new transaction with probability $q$ every second. We assume that there are $N=1000$ users. For each iteration, we simulate the network for 1000 seconds. Each plot is averaged over 100 iterations.

Once a user decides to start a transaction, a file size is chosen randomly in the following manner. We assume there are four types of files: $f_{doc}$ = 8KB (a document), $f_{image}$ = 1MB (an image), $f_{mp3}$ = 3 MB (a mp3 file), $f_{video}$ =  20 MB (a small video), and are chosen with probability $q_{doc}$, $q_{image}$, $q_{mp3}$, and $q_{video}$, respectively. In Figure \ref{fig:simulations_2}, we set $[q_{doc}, q_{image}, q_{mp3}, q_{video}] = [0.3, 0.3, 0.3, 0.1]$. This results in $\mu_f = 3.2$ MB as in Figure \ref{fig:analysis_2}. In Figure \ref{fig:simulations_1}, we set $[q_{doc}, q_{image}, q_{mp3}, q_{video}] = [0.26, 0.27, 0.27, 0.2]$, which gives $\mu_f = 5.08$ MB as in Figure \ref{fig:analysis_1}.

The simulation results show close concordance to our analysis. Note that the values in Figures \ref{fig:simulations_2} and \ref{fig:simulations_1} are slightly greater than that of Figures \ref{fig:analysis_2} and \ref{fig:analysis_1}. This is because, in the simulation, we round-up any fractional $N_{bs}$'s since the number of base stations needs to be integral.

\section{Analysis for the Number of Base Stations for TCP/NC and TCP}\label{sec:bs_tcpnc}

We now study the effect of TCP and TCP/NC's behavior (i.e. $\thru{} \leq \band$). We have shown that TCP/NC is robust against erasures; thus, allowing it to maintain a high throughput despite random losses. For example, if the network allows for 2 Mbps per user and there is 10\% loss rate, then the user should see approximately $2\cdot(1-0.1) = 1.8$ Mbps. Reference \cite{analysis} has shown, both analytically and with simulations, that TCP/NC indeed is able to achieve throughput close to 1.8 Mbps in such a scenario while TCP fails to do so.

We use the model and analysis from Sections \ref{sec:td}, \ref{sec:tcpnc}, and \ref{sec:simulations}. As in Section \ref{sec:simulations}, we set the maximum congestion window, $W_{\max}$, of TCP and TCP/NC to be 50 packets (with each packet being 1000 bytes long), and their initial window size to be 1. We consider $RTT$ = 100 ms and varying $p$ from 0 to 0.05. We note that, given $\band$ and $p$, $\thru{} \leq \band (1-p)$ regardless of the protocol used.

\begin{figure}[tbp]
\begin{center}\vspace*{.4cm}
\subfloat[$\thru{nc}$]{\includegraphics[width=.23\textwidth]{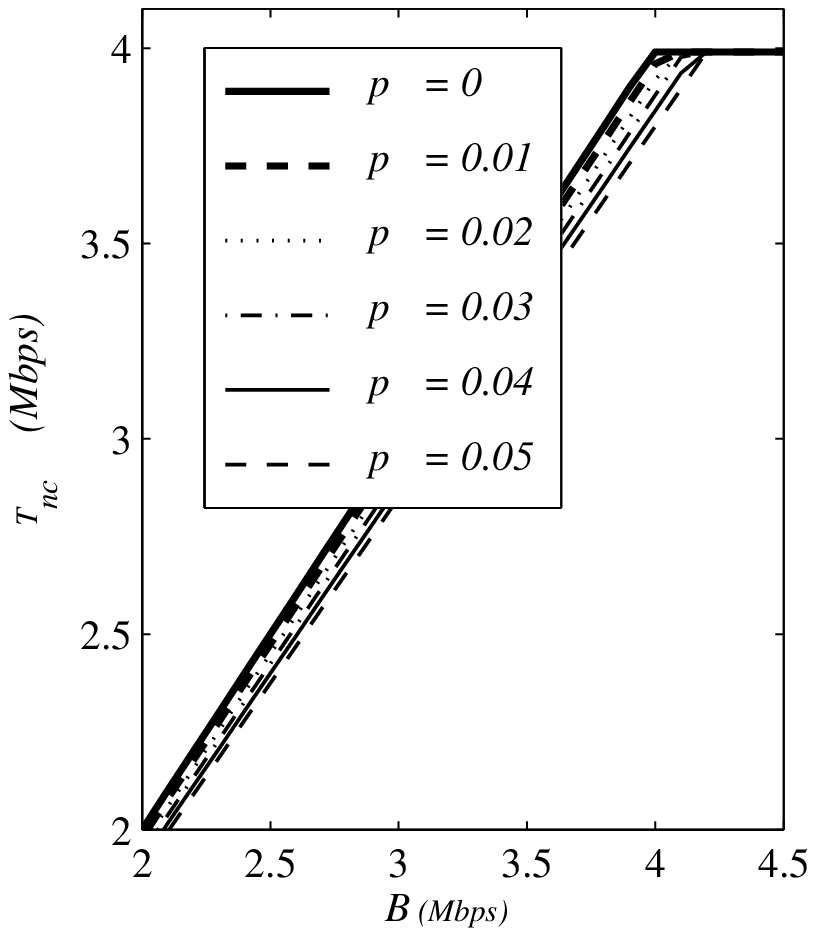}\label{fig:ncgoodput}}
\hspace*{.1cm}
\subfloat[$\thru{tcp}$]{\includegraphics[width=.226\textwidth]{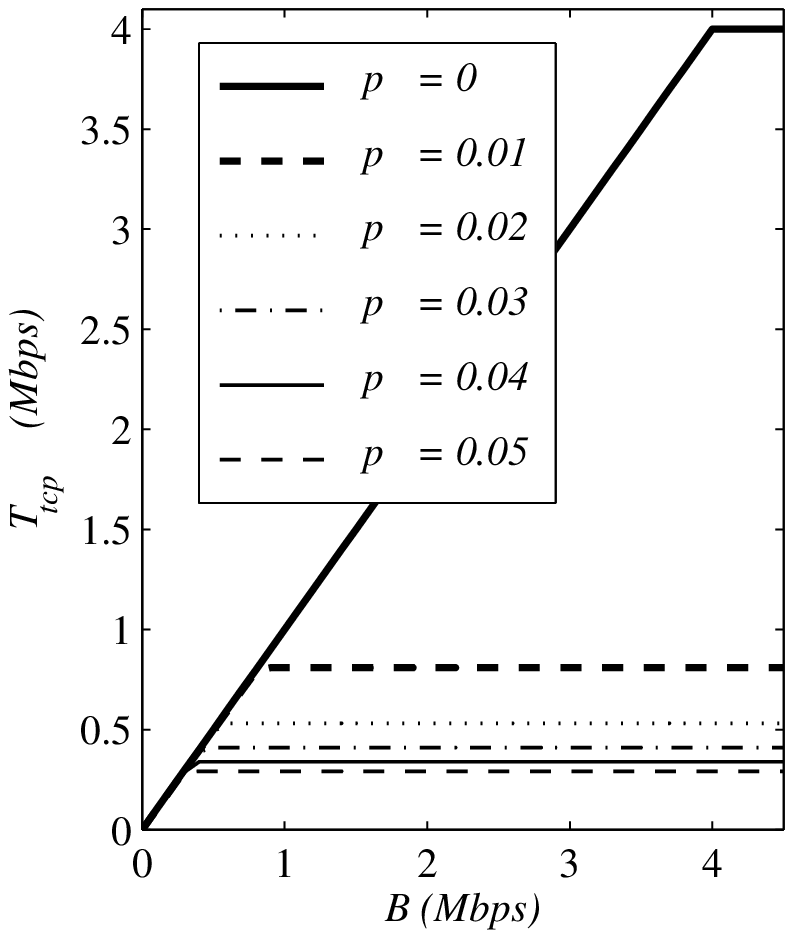}\label{fig:tcpgoodput}}
\end{center}\vspace*{-.2cm}
\caption{The value of $\thru{nc}$ and $\thru{tcp}$ against $\band$ for varying values of $p$. We set $RTT$ = 100 ms.}\label{fig:goodput}\vspace*{-.4cm}
\end{figure}


Combining Equation (\ref{eq:nc-final-longterm}) and $\thru{nc} \leq \band (1-p)$, we obtain the values of $\thru{nc}$ for various $\band$, $RTT$, and $p$. In Figure \ref{fig:ncgoodput}, the values of $\thru{nc}$ plateaus once $\band$ exceeds some value. This is caused by $W_{\max}$. Given $W_{\max}$ and $RTT$, TCP/NC and TCP both have a maximal throughput it can achieve. With the parameters we are considering, the maximal throughput is approximately 4 Mbps. Note that regardless of $p$, all TCP/NC flows achieve the maximal achievable rate. This shows that TCP/NC can overcome effectively the erasures or errors in the network, and provide a throughput that closely matches the bandwidth $\band$.


Combining Equation (\ref{eq:tcp}) and $\thru{tcp} \leq \band (1-p)$, we obtain the values of $\thru{tcp}$ for various $\band$, $RTT$, and $p$ as shown in Figure \ref{fig:tcpgoodput}. As in Figure \ref{fig:ncgoodput}, the values of $\thru{tcp}$ are also restricted by $W_{\max}$. However, TCP achieves this maximal throughput only when $p$ = 0. This is because, when there are losses in the network, TCP is unable to recover effectively from the erasures and fails to use the bandwidth dedicated to it. For $p > 0$, $\thru{tcp}$ is not limited by $W_{\max}$ but by TCP's performance limitations in lossy wireless networks.



\begin{figure*}[tbp]
\begin{center}
\subfloat[$p$ = 0]{\includegraphics[width=.22\textwidth]{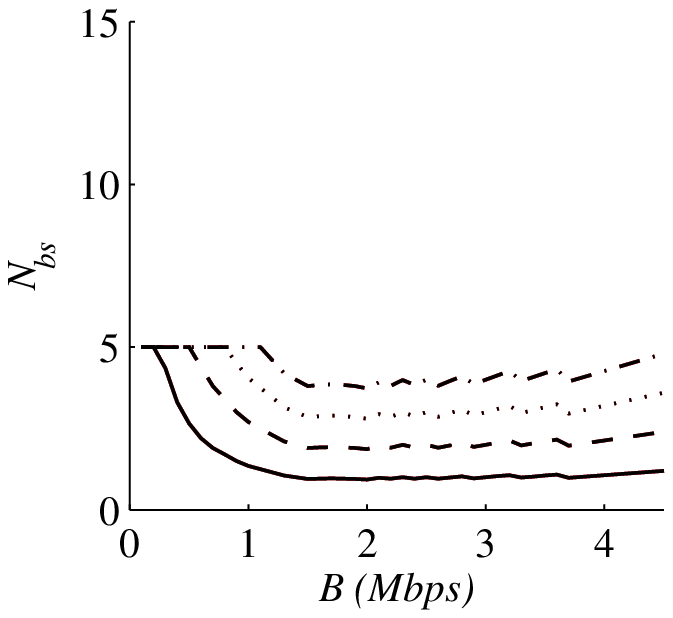}\label{fig:nbstcpnc100_320_0}}
\subfloat[$p$ = 0.01]{\includegraphics[width=.22\textwidth]{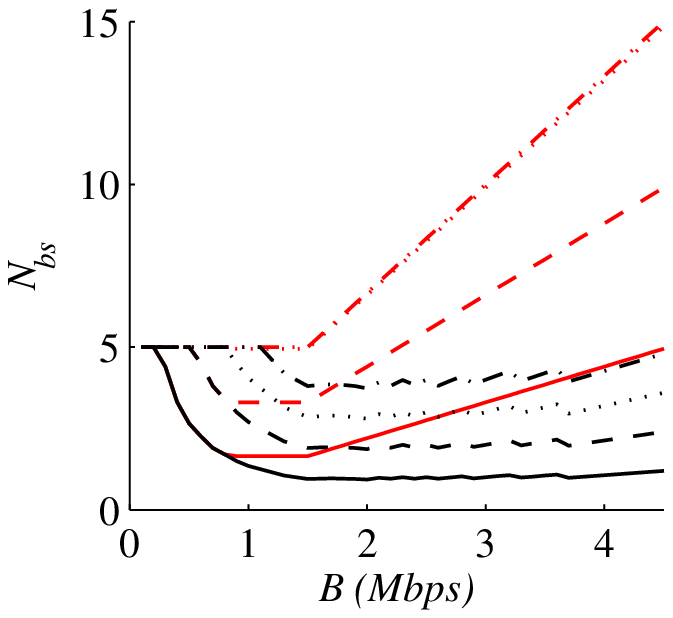}\label{fig:nbstcpnc100_320_1}}
\subfloat[$p$ = 0.02]{\includegraphics[width=.22\textwidth]{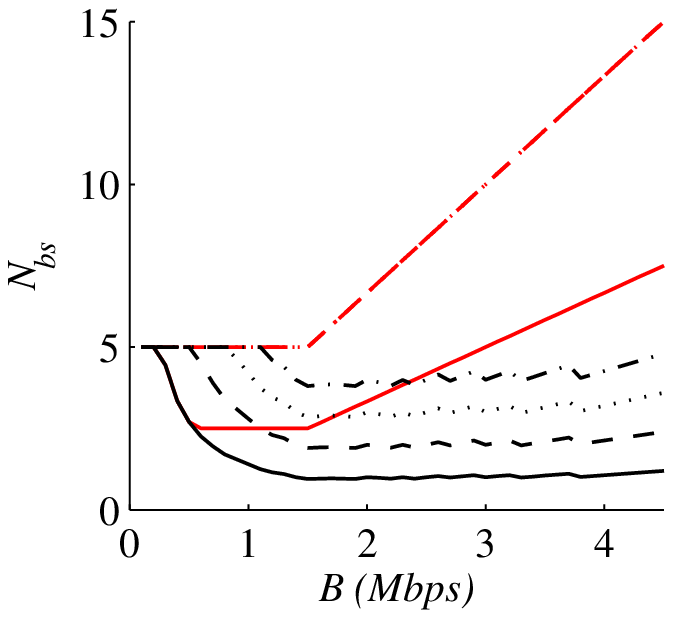}\label{fig:nbstcpnc100_320_2}}
\subfloat[$p$ = 0.05]{\includegraphics[width=.22\textwidth]{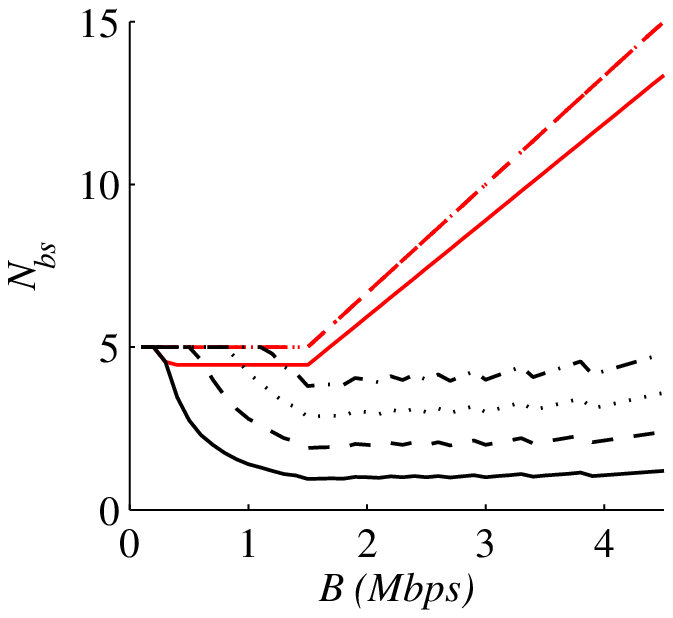}\label{fig:nbstcpnc100_320_5}}
\subfloat{\includegraphics[width=.12\textwidth]{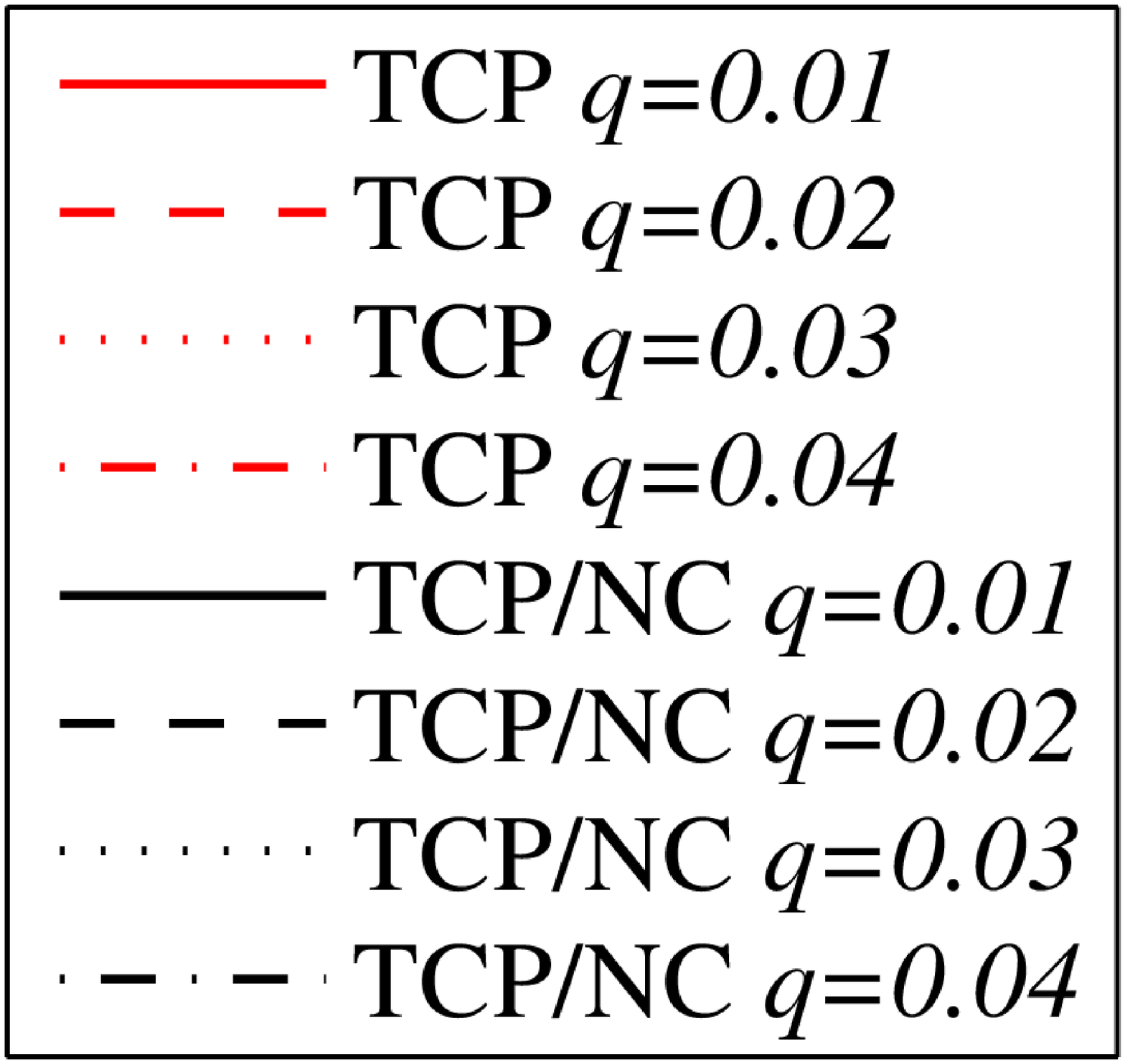}}
\end{center}\vspace*{-.2cm}
\caption{The value of $N_{bs}$ from Equation \eqref{eq:basestations} for TCP and TCP/NC with varying $p$ and $q$. Here, $RTT$ = 100 ms, $W_{\max} $ = 50, $N$ = 1000, and $\mu_f$ = 3.2 MB. In (a), $p = 0$ and both TCP and TCP/NC behaves the same; thus, the curves overlap. Note that this result is the same as that of Figure \ref{fig:analysis_2}. In (b), the value of $N_{bs}$ with TCP for $p = 0.03$ and 0.04 coincide (upper most red curve). In (c) and (d), the values of $N_{bs}$ with TCP for $p > 0.01$ overlap.}\label{fig:nbstcpnc100_320}\vspace*{-.4cm}
\end{figure*}
\begin{figure*}[tbp]
\begin{center}
\subfloat[$p$ = 0]{\includegraphics[width=.22\textwidth]{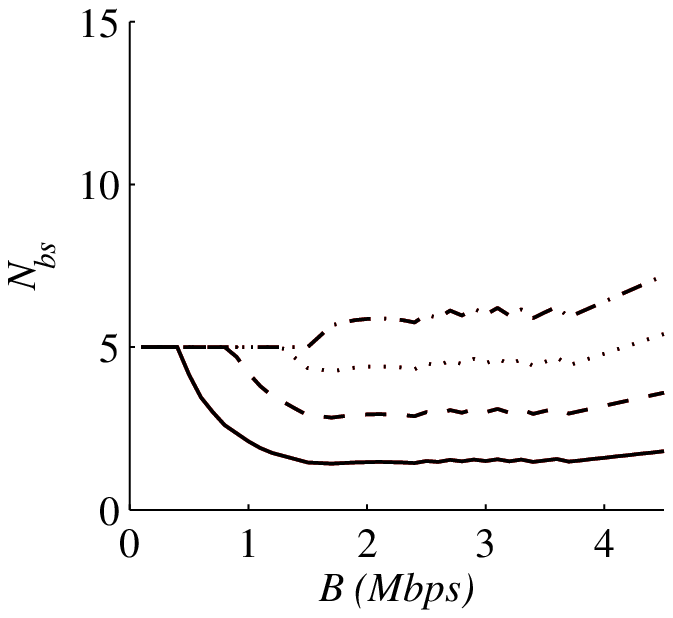}\label{fig:nbstcpnc100_580_0}}
\subfloat[$p$ = 0.01]{\includegraphics[width=.22\textwidth]{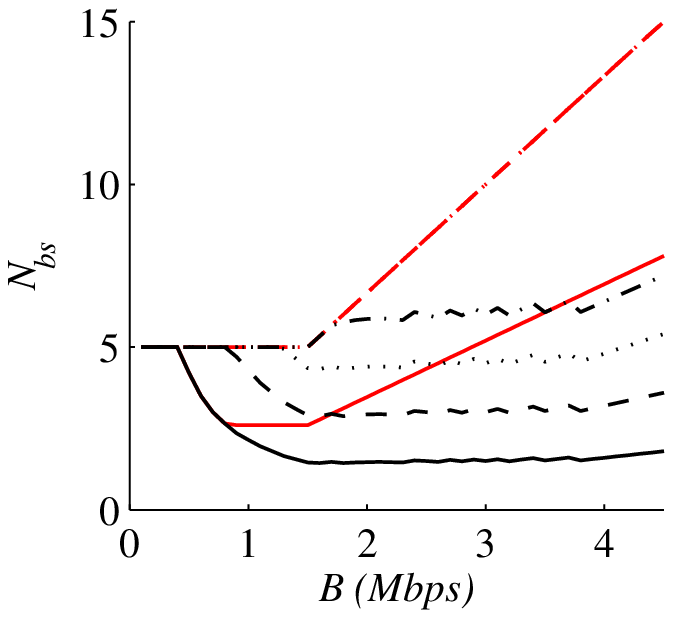}\label{fig:nbstcpnc100_580_1}}
\subfloat[$p$ = 0.02]{\includegraphics[width=.22\textwidth]{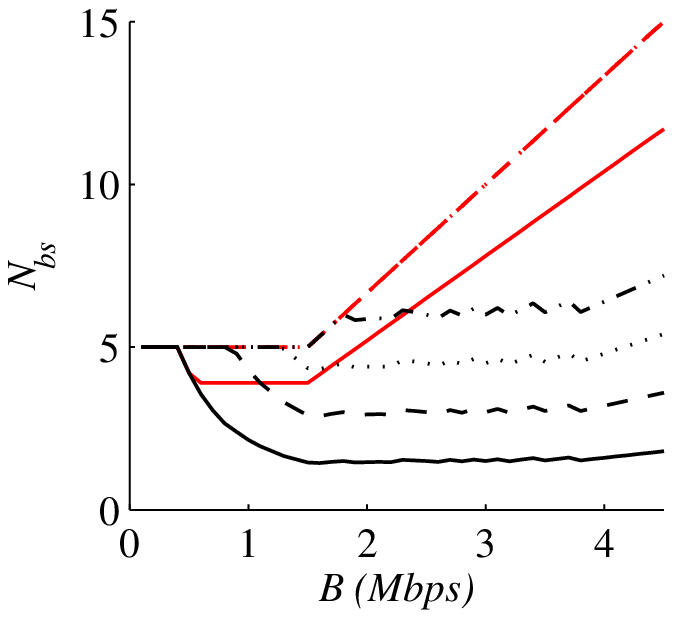}\label{fig:nbstcpnc100_580_2}}
\subfloat[$p$ = 0.03]{\includegraphics[width=.22\textwidth]{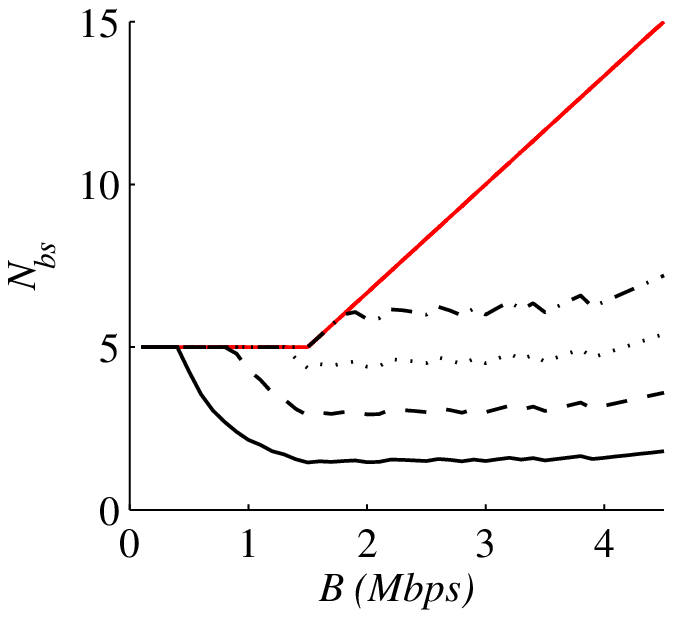}\label{fig:nbstcpnc100_580_3}}
\subfloat{\includegraphics[width=.12\textwidth]{legend_energy}}
\end{center}\vspace*{-.2cm}
\caption{The value of $N_{bs}$ from Equation \eqref{eq:basestations} for TCP and TCP/NC with varying $p$ and $q$. Here, $RTT$ = 100 ms, $W_{\max} $ = 50, $N$ = 1000, and $\mu_f$ = 5.08 MB. In (a), the results for TCP and TCP/NC are the same. Note that this result is the same as that of Figure \ref{fig:analysis_1}. In (b) and (c), the value of $N_{bs}$ with TCP for $p> 0.01$ coincide (upper red curve). In (d), the values of $N_{bs}$ with TCP for any $q$ all overlap. We do not show results for $p$ = 0.04 or 0.05 as they are similar to that of (d).}\label{fig:nbstcpnc100_508}\vspace*{-.2cm}
\end{figure*}

Using the results in Figure \ref{fig:goodput} and Equation (\ref{eq:basestations}), we can obtain the number of base stations $N_{bs}$ needed by both TCP and TCP/NC as shown in Figures \ref{fig:nbstcpnc100_320} and \ref{fig:nbstcpnc100_508}. TCP suffers performance degradation as $p$ increases; thus, $N_{bs}$ increases rapidly with $p$. Note that increasing $\band$ without being able to increase $\thru{}$ leads to inefficient use of the network, and this is clearly shown by the performance of TCP as $\band$ increases with non-zero loss probability, $p > 0$.

However, for TCP/NC, $N_{bs}$ does not increase significantly (if any at all) when $p$ increases. As discussed in Section \ref{sec:bs_analysis}, TCP/NC is able to translate better $\band$ into $\thru{nc}$ despite $p > 0$, i.e. $\band \approx \thru{nc}$. As a result, this leads to a significant reduction in $N_{bs}$ for TCP/NC compared to TCP. Note that $N_{bs}$ for TCP/NC is approximately equal to the values of $N_{bs}$ in Section \ref{sec:bs_analysis} regardless of the value of $p$. Since TCP/NC is resilient to losses, the behavior of $\thru{nc}$ does not change as dramatically against $p$ as that of $\thru{tcp}$ does. As a result, we observe $N_{bs}$ for TCP/NC to reflect closely the values of $N_{bs}$ seen in Section \ref{sec:bs_analysis}, which is the best case with $\band = \thru{}$.


We observe a similar behavior for other values of $RTT$ as we did for $RTT$ = 100 ms. The key effect of the value of $RTT$ in the maximal achievable throughput. For example, if $W_{\max}$ is limited to 50, the maximal achievable throughput is approximately 0.8 Mbps when $RTT$ = 500 ms, which is much less than the the 4 Mbps achievable with $RTT$ = 100 ms. As a result, for $RTT$ = 500 ms, neither $\thru{nc}$ nor $\thru{tcp}$ can benefit from the increase in $\band$ beyond 0.8 Mbps. Despite this limitation, TCP/NC still performs better than TCP when losses occur. When demand exceeds the maximal achievable throughput, $N_{bs}$ increases for both TCP/NC and TCP in the same manner. We do not present the results for want of space.

\section{Conclusions}\label{sec:conclusions}

We have presented an analytical study and compared the performance of TCP and TCP/NC. Our analysis characterizes the throughput of TCP and TCP/NC as a function of erasure probability, round-trip time, maximum window size, and the duration of the connection. We showed that network coding, which is robust against erasures and failures, can prevent TCP's performance degradation often observed in lossy networks. Our analytical model shows that TCP with network coding has significant throughput gains over TCP. TCP/NC is not only able to increase its window size faster but also to maintain a large window size despite losses within the network; on the other hand, TCP experiences window closing as losses are mistaken to be congestion. Furthermore, NS-2 simulations verify our analysis on TCP's and TCP/NC's performance. Our analysis and simulation results both support that TCP/NC is robust against erasures and failures. Thus, TCP/NC is well suited for reliable communication in lossy wireless networks.

In addition, we studied the number of base stations $N_{bs}$ needed to improve the throughput to the users. It may seem that higher throughput necessarily increases $N_{bs}$. Indeed, if there are enough demand (i.e. high throughput per connection, many active users in the network, etc.), we eventually need to increase $N_{bs}$. However, we show that this relationship is not necessarily true. When the observed throughput by the user is low, each transaction takes more time to complete and each user stays in the system longer. This degrades the user experience and delays the release of network resources dedicated to the user. This is particularly important as the number of active users each base station can support is limited to the low hundreds. We observed that, given bandwidth allocated a user, achieving low throughput may lead to a significant increase in $N_{bs}$ and an ineffective use of the network resources; while achieving high throughput may lead to reduction in $N_{bs}$. We showed that TCP/NC, which is more resilient to losses than TCP, may better translate bandwidth to throughput. Therefore, TCP/NC may lead to a better use of the available network resources and reduce the number of base stations $N_{bs}$ needed to support users at a given throughput.

\bibliography{References}
\bibliographystyle{IEEEtran}

\end{document}